\documentclass[]{ar2e-changed}

\usepackage{amsmath,amssymb}
\usepackage{pifont}
\usepackage{latexsym}
\usepackage{graphicx}
\usepackage{dcolumn}
\usepackage{bm}
\usepackage{psfrag}
\usepackage{subfigure}
\usepackage{tabularx}
\usepackage{verbatim}
\usepackage{placeins}

\def\ek{\varepsilon_{\bf k}}
\def\Gapat{\Delta_{\mathrm{at}}}
\def\Ubar{\,\mathcal{U}}
\def\Up{U^\prime}
\def\Ueff{U_{\mathrm{eff}}}
\def\Wt{\widetilde{W}}
\def\hn{\hat{n}}
\def\hN{\hat{N}}
\def\hnmu{\hat{n}_{m\uparrow}}
\def\hnmd{\hat{n}_{m\downarrow}}

\def\hnpmd{\hat{n}_{m^\prime\downarrow}}
\def\spinup{\uparrow}
\def\spindown{\downarrow}

\def\ket{\rangle}
\def\geff{g_{\mathrm{eff}}}
\def\chiloc{\chi_{\mathrm{loc}}}
\def\TFL{T_{\mathrm{FL}}}
\def\chiloc{\chi_{\mathrm{loc}}}
\def\Seff{S_{\mathrm{eff}}}
\def\mloc{m_{\mathrm{loc}}}

\def\vk{{\bf k}}

\def\vr{{\bf r}}
\def\vrp{{\bf r}^\prime}

\def\eV{\,\mbox{eV}\,}
\def\meV{\,\mbox{meV}\,}

\def\mp{m^\prime}
\def\mpp{m^{\prime\prime}}
\def\ttg{$t_{2g}\,$}
\def\eg{$e_{g}\,$}

\def\dx2{$d_{x^2-y^2}$}
\def\dz2{$d_{z^2}$}

\def\HEM{$\alpha$-Fe$_2$O$_3\,$}
\def\LCO{LaCoO$_3\,$}
\def\LiVO{LiV$_2$O$_4$\,}
\def\BaVS{BaVS$_3$\,}
\def\SVO{SrVO$_3$\,}
\def\CVO{CaVO$_3$\,}
\def\LTO{LaTiO$_3$\,}
\def\YTO{YTiO$_3$\,}
\def\SCO{SrCrO$_3$\,}
\def\SMO{SrMnO$_3$\,}

\def\SRO{Sr$_{}$Ru$_{}$O$_{3}$\,}
\def\SROtwo{Sr$_{2}$Ru$_{}$O$_{4}$\,}

\def\CRO{Ca$_{}$Ru$_{}$O$_{3}$\,}
\def\CROtwo{ Ca$_{2}$Ru$_{}$O$_{4}$\,}

\def\CSROtwo{Ca$_{2-x}$Sr$_{x}$Ru$_{}$O$_{4}$\,}

\def\BFA{BaFe$_2$As$_2$}

\begin{document}

\title{Strong 
correlations from Hund's coupling}
\markboth{Georges, de' Medici\& Mravlje}{Hund's Correlated Materials}

\author{Antoine Georges
\affiliation{
Coll\`ege de France, 11 place Marcelin Berthelot, 75005 Paris, France
\\
Centre de Physique Th\'eorique, \'Ecole Polytechnique, CNRS, 91128 Palaiseau Cedex, France
\\
DPMC-MaNEP, Universit\'e de Gen\`eve, 24 quai Ernest Ansermet, CH-1211 Gen\`eve, Suisse 
}
Luca de' Medici
\affiliation{Laboratoire de Physique des Solides, UMR8502 CNRS-Universit\'e Paris-Sud, Orsay, France}
Jernej Mravlje
\affiliation{
Centre de Physique Th\'eorique, \'Ecole Polytechnique, CNRS, 91128 Palaiseau Cedex, France
\\
Coll\`ege de France, 11 place Marcelin Berthelot, 75005 Paris, France
\\
Jo\v{z}ef Stefan Institute, Jamova~39, SI-1000, Ljubljana, Slovenia}
}


\begin{abstract}
Strong electronic correlations are often associated with the proximity of a Mott insulating state. 
In recent years however, it has become increasingly clear that the  
Hund's rule coupling (intra-atomic exchange) is responsible for 
strong correlations in multi-orbital metallic materials which are not close 
to a Mott insulator. 
Hund's coupling has two effects: it influences the energetics of the Mott gap and strongly 
suppresses the coherence scale for the formation of a Fermi-liquid. 
A global picture has emerged recently, which emphasizes the importance of the 
average occupancy of the shell as a control parameter.  
The most dramatic effects occur away from half-filling or single occupancy. 
The theoretical understanding and physical properties of these `Hund's metals' are reviewed, together 
with the relevance of this concept to transition-metal oxides of the $3d$, and especially $4d$ series (such as ruthenates), 
as well as to the iron-based superconductors (iron pnictides and chalcogenides).
\end{abstract}
\date{\today}
\maketitle

\section{Introduction}
\label{sec:intro}

The electronic state of many materials with partially filled d- or f-shells, as well as molecular solids, 
is characterized by strong correlations. Picturing their wave-function as a determinant of 
single-particle states does not properly account for their physical properties.  
Materials with strong electronic correlations display fascinating phenomena, often with a large amplitude, such as metal-insulator 
transitions, high-temperature superconductivity, colossal magnetoresistance, a large thermoelectric power, or 
carriers with a large effective mass and reduced spectral weight.  
%

The Mott phenomenon - the localization of electrons due to the strong Coulomb repulsion and the 
reduced bandwidth - has emerged as a central paradigm in this field~\cite{imada_mit_review}. 
The parent compounds of high-temperature cuprate superconductors are widely considered to be 
Mott insulators (of the so-called `charge-transfer' type). The metallic and superconducting states 
emerge by doping this insulator with charge carriers. 
In this view, strong electronic correlations in the metallic state are due to the proximity of the Mott 
insulator. Hence `Mottness' is widely regarded as being key to the strong correlations observed in 
oxides and organic compounds. 
%

Cuprates have a single active electronic band at the Fermi-level, a rather unique property which incidentally 
is also shared by the organic superconductors of the BEDT family.  
With very few exceptions, known oxides of other transition metals are in contrast multi-band materials, and so 
are the recently discovered iron-based superconductors. Several bands cross the Fermi level, formed by the 
different orbitals of the transition-metal $d$-shell hybridizing with ligands~\cite{TokuraNagaosa_OrbitalPhysics}.  
Many of these multi-orbital materials, such as ruthenates and iron pnictides and chalcogenides, 
are metals which display clear signatures of strong correlations while not being close to a Mott insulating 
state.
%
This raises a puzzling question: what is the physical origin of electronic correlations in these 
materials ?

In the last few years, there has been increasing awareness that Hund's coupling may be responsible 
for these effects. 
Hund's coupling is the energy scale associated with intra-atomic exchange, which lowers the cost in repulsive 
Coulomb energy when placing two electrons in different orbitals with parallel spin, as opposed to two 
electrons in the same orbital~\cite{hund_1925}. 
This shakes the paradigm establishing `Mottness' as the unique origin of strong correlations, and 
highlights that another class of strongly correlated but itinerant systems 
have physical properties distinctly different from doped Mott insulators.
The term `Hund's metals' has been coined in Ref.~\cite{Yin_kinetic_frustration_allFeSC} to designate such materials. 

There are two distinct effects of the Hund's rule coupling. 
The first is a high-energy effect. As emphasized early on~\cite{vandermarel_sawatzky_prb_1988,vandermarel_phd_1985}, 
the effective Coulomb repulsion for an isolated atom is increased by Hund's coupling for a half-filled shell, 
while it is decreased for all other fillings.
The second is a low-energy effect, revealed in early studies of a single impurity atom coupled to a 
conduction electron gas (the Kondo problem). For a multi-orbital shell, the characteristic temperature 
below which screening of the atomic degrees of freedom takes place is considerably lowered 
by Hund's coupling~\cite{okada73,jayaprakash81,jones87,kusunose97,yotsuhashi01,pruschke_epjb_2005,nevidomskiy09}.
This is due to the quenching of orbital momentum and associated loss of exchange energy, and 
explains the sensitive dependence of the Kondo temperature 
on the size of the impurity spin~\cite{schrieffer_japplphys_1967,blandin_japplphys_1968,daybell_rmp_1968}. 

What is remarkable is that these effects, documented for an isolated
atom or for a single atomic impurity in a metallic host, continue to
play a crucial role in the context of itinerant systems with a
bandwidth significantly larger than Hund's coupling.
That this is the case has been demonstrated in several recent studies, which led to the realization that `Hundness' 
is the key explanation of electronic correlations in several families of metallic systems. 
Two remarkable theoretical studies, in the context of a 5-band description 
of the metallic state of iron pnictides~\cite{haule09} 
and in that of a 3-band multi-orbital Hubbard-Kanamori model~\cite{werner08}, revealed that the 
low-energy quasiparticle coherence scale is considerably reduced by Hund's coupling. 
Such a reduction was also emphasized in the context of ruthenates in Ref.~\cite{mravlje11}.
This leaves an incoherent metallic state with frozen local moments in an extended temperature range above the coherence scale, 
for which the authors of Ref.~\cite{werner08} coined the term `spin-freezing' regime. 
These authors also discovered that this regime displays non-Fermi liquid properties of the self-energy, 
characterized by a power-law behaviour. 

\begin{figure}[!ht]
\begin{center}
\includegraphics[width=15cm]{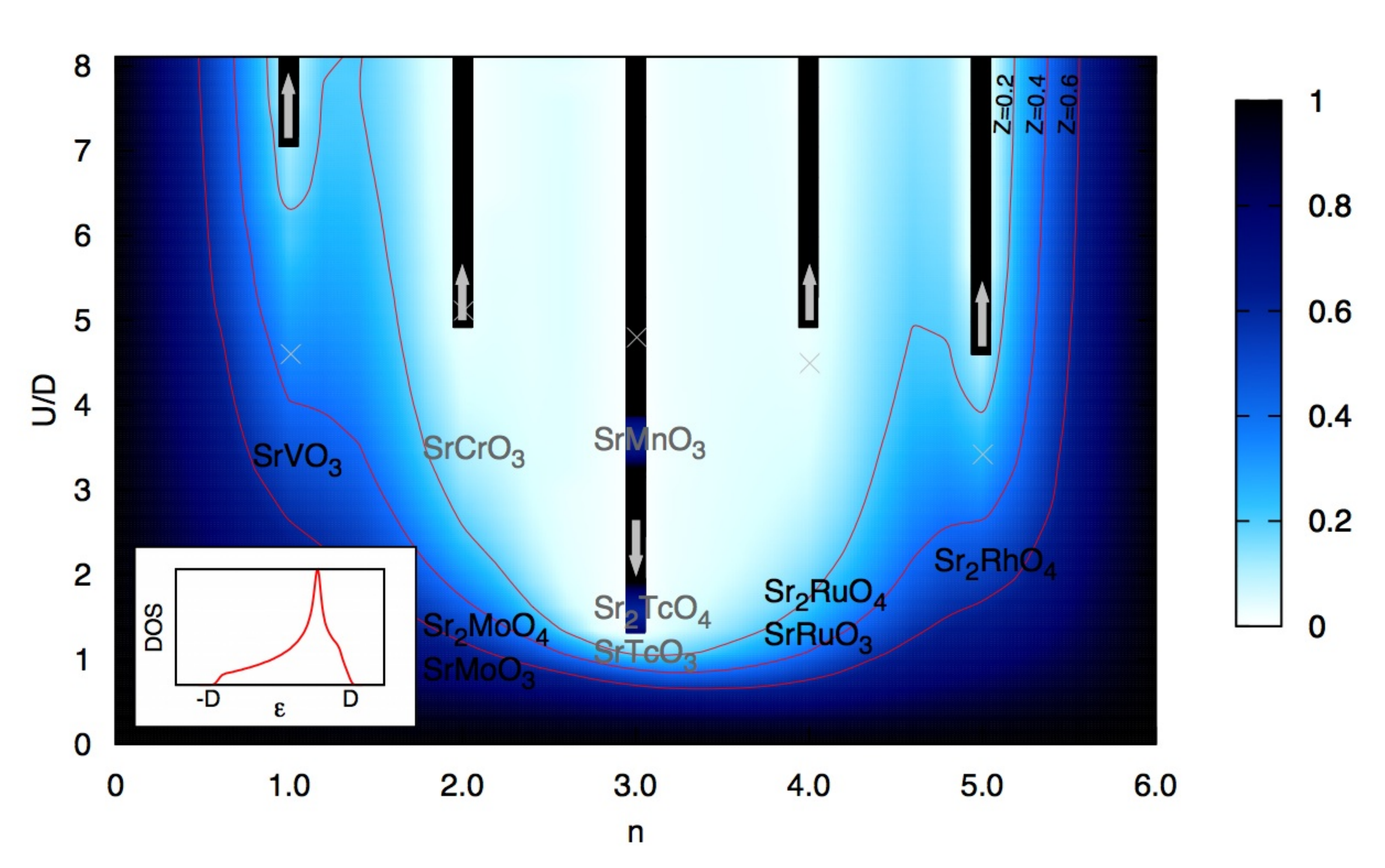}
\end{center}
\caption{Colour intensity map of the `degree of correlation' (as measured by the 
quasiparticle weight $Z$ - right scale) for a Hubbard-Kanamori model with 3 orbitals 
appropriate to the description of early transition-metal oxides with a partially occupied \ttg shell.  
The vertical axis is the interaction strength $U$ normalized to the half-bandwidth $D$, and a finite 
Hund's coupling $J=0.15\,U$ is taken into account.   
The horizontal axis is the number of electrons per site - from 0 (empty shell) to 6 (full shell).  
Darker regions correspond to good metals and lighter regions to correlated metals. 
The black bars signal the Mott-insulating phases for $U>U_c$. 
The arrows indicate the evolution of $U_c$ upon further increasing $J$, and emphasize the 
opposite trend between half-filling and a generic filling. Crosses denote the values of $U_c$ for $J=0$. 
  One notes that, among integer fillings, the case of 2 electrons (2 holes) displays correlated behaviour 
  in an extended range of coupling, with `spin-freezing' above some low coherence scale.   
  Specific materials are schematically placed on the diagram. 
  The materials denoted in black have been placed according to the experimental value of 
  $\gamma/\gamma_\mathrm{LDA}$. 
For detailed explanations, see Sec.~\ref{sec:janus}. 
The DMFT calculations leading to a related plot in Ref.~\cite{demedici_prl_2011} have been 
repeated here using a more realistic DOS for \ttg states (inset).  
\label{fig:Zcontour}}
\end{figure}

As shown in Refs.~\cite{werner09,demedici_MottHund},  
the influence of Hund's coupling on the  energetics of charge-transfer in an isolated atom 
has important consequences for the Mott critical coupling in a solid. 
The generic effect (when orbital degeneracy is preserved) 
is that non half-filled materials are driven further away from the Mott insulating state. 
In Ref.~\cite{demedici_prl_2011}, a global picture was proposed, which also shows how to place many different 
materials on a map parametrized by the interaction strength and the filling of the shell. 
It was emphasized there that the two key effects  
compete with one another in the generic case of a non half-filled shell:  Hund's coupling 
drives the system away from the Mott transition but at the same time makes the metallic state 
more correlated by lowering the quasiparticle coherence scale. Like the Roman god Janus, the 
Hund's rule coupling has two faces !
This global picture is illustrated and summarized on Fig.~\ref{fig:Zcontour} (which is discussed 
in much greater details in Sec.~\ref{sec:janus}). 

At a fundamental level, a key lesson is that intra-atomic correlations play a crucial role even 
in itinerant systems with relatively broad bands and moderate Hubbard repulsion, 
such as transition-metals of the 4d and 5d series or iron pnictides and chalcogenides. 
Dynamical mean-field theory (DMFT)~\cite{georges_review_dmft,kotliar_dmft_physicstoday} 
is currently the most appropriate theoretical framework to deal 
with these issues, since it handles band-like and atomic-like aspects on equal footing. 
In contrast to more conventional approaches picturing a solid as an inhomogeneous electron gas to which interactions 
are added perturbatively, DMFT emphasizes local many-body correlations by viewing 
a solid as an ensemble of self-consistently hybridized atoms.  

This article is organized as follows. 
In Sec.~\ref{sec:hund}, an introduction to Hund's rules and intra-atomic Coulomb 
interactions in the multi-orbital context is provided. 
In Sec.~\ref{sec:mottgap} the influence of Hund's coupling on the intra-atomic charge gap and 
the Mott critical coupling is explained. 
Sec.~\ref{sec:kondo} reviews the influence of Hund's coupling on the Kondo temperature of 
a multi-orbital impurity atom in a metallic host. 
Sec.~\ref{sec:dmft} briefly introduces dynamical mean-field theory, which provides a bridge 
between single-atom physics and the full solid. 
Sec.~\ref{sec:janus} is the core part of this article, in which the key effects of the Hund's rule coupling in the 
solid-state context are put together. 
Sec.~\ref{sec:ruthenates} and Sec.~\ref{sec:pnictides} consider ruthenates and 
iron pnictides/chalcogenides, respectively, in the perspective of Hund's metals.

\section{Intra-atomic exchange and the Hund's rule coupling}
\label{sec:hund}

In 1925, in an article dealing with the spectra of transition-metal atoms~\cite{hund_1925}, 
Friedrich Hund formulated a set of rules specifying the ground-state configuration of 
multi-electron atomic shells. 
For $N$ electrons in a shell with orbital degeneracy $M$ ($=2l+1$), the rules state that: 
\begin{itemize}
\item Total spin $S$ should first be maximized (rule of `maximum multiplicity') 
\item Given $S$, total angular momentum $L$ should be maximized 
\item Finally, the lowest $J=|L-S|$ should be selected for $N<M$ (less than half-filled shell) and the 
highest $J=L+S$ for $N>M$. 
\end{itemize}
For example a d-shell with 3 electrons will have $S=3/2, L=3, J=3/2$ (e.g.  
$\uparrow,\uparrow,\uparrow,0,0$), with 6 electrons 
$S=L=2, J=4$ (e.g. $\uparrow\downarrow,\uparrow,\uparrow,\uparrow,\uparrow$), while 
the half-filled shell with 5 electrons (e.g. $\uparrow,\uparrow,\uparrow,\uparrow,\uparrow$)  
has $S=J=5/2$ and a fully quenched angular momentum $L=0$. 
These rules are sometimes referred to as the `bus-seat' rule: 
singly-occupied spots are filled first, then double occupancies are created when singly-occupied 
spots are no longer available.  

The origin of these rules is traditionally attributed to the minimization of the Coulomb interaction between electrons. 
For two electrons for example, the first rule (S=1 rather than S=0) forces an antisymmetric wave-function of the radial part, 
so that `electrons are further apart'. 
In quantum-mechanical terms, the energy gain associated with Hund's
rule is the intra-atomic exchange energy~\footnote{Besides such a gain which
determines the ordering of multiplets in calculations where the
single-electron basis is fixed~\cite{slater_phys_rev_1929}, another term
appears in self-consistent calculations in which the single-orbital basis is
allowed to vary~\cite{levine_quantum_chemistry_book}. This other term, which comes from the smaller
screening of the electron-nucleus interaction for high-spin and
high-orbital momentum states~\cite{boyd_hund_nature_1984}, becomes
dominant for light neutral atoms. For a recent discussion and references to further reading, see Ref.~\cite{oyamada_j_chem_phys_2010}.}.
The third rule is associated with spin-orbit coupling,  
which we shall not consider in this paper although its physical effects have attracted considerable attention recently.

To illustrate these rules in a more quantitative form appropriate to the solid-state context of this article, 
let us consider the hamiltonian describing 
the \ttg triplet of orbitals with lowest energy of a transition-metal ion in a cubic crystal field with an octahedral environment. 
The case of two orbitals and an \eg doublet is considered in details in Appendix~\ref{sec:appendix}. 
For both \eg and \ttg, there are only three independent Coulomb integrals, which are matrix elements of the {\it screened} Coulomb interaction 
in appropriately chosen wavefunctions of the \ttg orbitals in the solid:
\begin{eqnarray}
U\,&=&\, \int d\vr d\vrp\, |\phi_m(\vr)|^2\, V_c(\vr,\vrp)\, |\phi_m(\vrp)|^2 
\nonumber \\
U^\prime\,&=&\, \int d\vr d\vrp\, |\phi_m(\vr)|^2\, V_c(\vr,\vrp)\, |\phi_{\mp}(\vrp)|^2 
\nonumber \\
J\,&=&\, \int d\vr d\vrp\, \phi_m(\vr)\phi_{\mp}(\vr)\, V_c(\vr,\vrp)\, \phi_m(\vrp)\phi_{\mp}(\vrp) 
\label{eq:uupj}
\end{eqnarray}
Indeed, the wave functions can be chosen real (so that the `spin-exchange' and 
`pair-hopping' integrals are equal $J=J^\prime$),  
and all other terms in the interaction tensor, e.g. of the type $U_{mmm\mp}$ vanish by symmetry in this case. 
Because there are no other exchange integrals involved, the full many-body atomic hamiltonian for 
\ttg states takes the Kanamori form~\cite{kanamori_ptp_1963} :
\begin{eqnarray}
H_{\mathrm{K}}\,=\,U\sum_m \hnmu\hnmd\,+\,U^\prime\sum_{m\neq\mp} \hnmu\hnpmd\,
+(U^\prime-J) \sum_{m<\mp,\sigma} \hn_{m\sigma}\hn_{\mp\sigma} + \nonumber \\
-J\,\sum_{m\neq\mp} d^+_{m\spinup}d_{m\spindown}\,d^+_{\mp\spindown}d_{\mp\spinup} \,
+ J\, \sum_{m\neq\mp} d^+_{m\spinup}d^+_{m\spindown}\,d_{\mp\spindown}d_{\mp\spinup} 
\label{eq:ham_kanamori} 
\end{eqnarray}
The first three terms involve only density-density interactions, between electrons with opposite spins in the 
same orbital ($U$), opposite spins in different orbitals ($\Up<U$) and parallel spins in different orbitals. The latter case 
has the smallest coupling $\Up-J$, reflecting Hund's first rule. 
%

For later use, it will be useful to consider a generalization of this Kanamori multi-orbital hamiltonian to a form in which 
all coupling constants are independent:
\begin{eqnarray}
H_{\mathrm{GK}}\,=\,U\sum_m \hnmu\hnmd\,+\,U^\prime\sum_{m\neq\mp} \hnmu\hnpmd\,
+(U^\prime-J) \sum_{m<\mp,\sigma} \hn_{m\sigma}\hn_{\mp\sigma} + \nonumber \\
-J_X\,\sum_{m\neq\mp} d^+_{m\spinup}d_{m\spindown}\,d^+_{\mp\spindown}d_{\mp\spinup} \,
+ J_P\, \sum_{m\neq\mp} d^+_{m\spinup}d^+_{m\spindown}\,d_{\mp\spindown}d_{\mp\spinup} 
\label{eq:ham_kanamori_general} 
\end{eqnarray}
Defining the total charge, spin and orbital isospin generators ($\vec{\tau}$ are the Pauli matrices): 
\begin{equation}
\hat{N} = \sum_{m\sigma} \hn_{m\sigma}\,\,\,,\,\,\,
\vec{S} = \frac{1}{2} \sum_m \sum_{\sigma\sigma^\prime} 
d^\dagger_{m\sigma} \vec{\tau}_{\sigma\sigma^\prime}d_{m\sigma^\prime}\,\,\,,\,\,
L_m = i\sum_{\mp\mpp}\sum_\sigma\epsilon_{m\mp\mpp} d^\dagger_{\mp\sigma}d_{\mpp\sigma},
\end{equation}
%
the generalized Kanamori hamiltonian (\ref{eq:ham_kanamori_general}) can be rewritten as:
\begin{eqnarray}
&H_{\mathrm{GK}}=\frac{1}{4} (3\Up-U) \hN(\hN-1)+(\Up-U)\vec{S}^2+\frac{1}{2}(\Up-U+J)\vec{L}^2 
+(\frac{7}{4}U-\frac{7}{4}\Up - J)\hat{N} + \nonumber \\
&+(\Up-U+J+J_P)\,\sum_{m\neq\mp} d^+_{m\spinup}d^+_{m\spindown}\,d_{\mp\spindown}d_{\mp\spinup}
+(J-J_X)\sum_{m\neq\mp} d^+_{m\spinup}d_{m\spindown}\,d^+_{\mp\spindown}d_{\mp\spinup}
\label{eq:ham_t2g_general_NSL}
\end{eqnarray}
It thus has full $U(1)_C\otimes SU(2)_S\otimes SO(3)_O$ symmetry provided  $J_X=J$ and $J_P=U-U^\prime-J$, 
in which case the hamiltonian reduces to the first line in Eq.~(\ref{eq:ham_t2g_general_NSL}). 
We shall loosely refer to such symmetry as `rotational invariance'. 
Note that rotational invariance of $H_{\mathrm{GK}}$ does {\it not} imply that $\Up$ and $U$ are related. 
In particular for $J_X=J$ and $\Up=U-J$ ($J_P=0$), one obtains a minimal rotationally-invariant hamiltonian  
$(U-3J/2)\hN(\hN-1)/2-J\vec{S}^2$ involving only $\hN^2$ and $\vec{S}^2$, to be discussed in more details below 
(Eqs.~(\ref{eq:ham_DN}) and (\ref{eq:ham_dworin_2orbital})). This actually holds for an arbitrary number $M$ of orbitals.  

Using (\ref{eq:ham_t2g_general_NSL}), the physical \ttg hamiltonian (\ref{eq:ham_kanamori}) which has $J_X=J_P=J$ 
is seen to be rotationally invariant provided: 
\begin{equation}
\Up\,=\,U-2J
\label{eq:up_u}
\end{equation}
in which case the hamiltonian takes the form:
\begin{equation}
H_{t_{2g}}\,=\,(U-3J)\,\frac{\hN(\hN-1)}{2}\,-\,2J\,\vec{S}^2-\frac{J}{2}\vec{L}^2+\frac{5}{2}J\,\hat{N}
\label{eq:ham_t2g_NSL}
\end{equation}
In this form, Hund's first two rules (maximal $S$, then maximal $L$) are evident. 
The spectrum of this hamiltonian is detailed in Table \ref{table:spectrum_t2g}. 
%
\begin{table}[!b]
\begin{center}
\begin{tabular}{c c c c c}
N & S & L & Degeneracy $=(2S+1)(2L+1)$ & Energy \\ 
\hline\hline
0,[6] & 0 & 0 & \boxed{1} & 0 \\ \hline
1,[5] & $1/2$ & 1 & \boxed{6} & $-5J/2$,[$10\Ubar-5J/2$]  \\ \hline
2,[4] & $1$ & 1 & \boxed{9} & $\Ubar-5J$,[$6\Ubar-5J$]  \\ 
2,[4] & $0$ & 2 & 5 & $\Ubar-3J$,[$6\Ubar-3J$] \\
2,[4] & $0$ & 0 & 1 & $\Ubar$,[$6\Ubar$] \\ \hline
3 & $3/2$ & 0 & \boxed{4} & $3\Ubar-15 J/2 $ \\
3 & $1/2$ & 2 & 10 & $3\Ubar-9J/2$ \\
3 & $1/2$ & 1 & 6 & $3\Ubar-5J/2$ \\
\end{tabular}
\end{center}
\caption{Eigenstates and eigenvalues of the \ttg Hamiltonian 
$\Ubar \hN(\hN-1)/2-2J\vec{S}^2-J\vec{L}^2/2$
in the atomic limit ($\Ubar\equiv U-3J$).  
The boxed numbers identifies the ground-state multiplet and its degeneracy, for $J>0$.
\label{table:spectrum_t2g}} 
\end{table}
%

Condition (\ref{eq:up_u}) is realized if $U,\Up,J$ are calculated assuming a spherically symmetric interaction and the 
\ttg wave-functions resulting from simple crystal-field theory. In this approximation, these integrals can be 
expressed in terms of Slater parameters $F^0,F^2,F^4$ (or alternatively Racah parameters $A,B,C$)~\cite{sugano_multiplets_book} : 
\begin{eqnarray}
U&=&F^0+\frac{4}{49}F^2+\frac{4}{49}F^4=A+4B+3C \nonumber\\
U^\prime&=&F^0-\frac{2}{49}F^2-\frac{4}{441}F^4=A-2B+C=U-2J\nonumber\\
J&=&\frac{3}{49}F^2+\frac{20}{441}F^4=3B+C
\label{eq:slater_t2g}
\end{eqnarray}
A rotationally invariant form of the \ttg hamiltonian is obtained when assuming spherical symmetry 
because the orbital angular momentum in the \ttg states is 
only partially quenched, from $l=2$ down to $l=1$. The orbital isospin generators are thus closely related 
to those of angular momentum with $l=1$ (up to a sign, cf.~\cite{sugano_multiplets_book}). 
In the solid-state, $V_c$ is the {\it screened} Coulomb interaction. The spherical symmetry of $V_c$ 
is of course no longer exact, but often considered to be a reasonable approximation so that 
$U^\prime=U-2J$ is often used in the solid as well. 

For an entire d-shell, the Kanamori hamiltonian~(\ref{eq:ham_kanamori}) is not exact and a full 
interaction tensor $U_{m_1m_2m_3m_4}$ must be considered. For an isolated atom with spherical 
symetry, this tensor can be parametrized in terms of three independent Slater (Racah) parameters 
$F^0,F^2,F^4$, while $9$ parameters are needed in principle in cubic 
symmetry~\cite{sugano_multiplets_book}. 
%
%
%
A word of caution is in order regarding notations. For an entire d-shell, 
it is customary~\cite{vandermarel_sawatzky_prb_1988,vandermarel_phd_1985,haverkort_phd_2005} 
to define 
$U_d=F^0$, the Hund's rule coupling $J_H=(F^2+F^4)/14$ and a third parameter 
$14C_d=9F^2/7-5F^4/7$. Those should not be confused with the $U$ and $J$ couplings defined 
above for a \ttg and \eg shell. For example, using (\ref{eq:slater_t2g},\ref{eq:slater_eg}) 
$U=F^0+8 J_H/7$, 
$J=5J_H/7+C_d/9 $ for \ttg and $J=30 J_H/49+4C_d/21$ for \eg.

We lack space here to discuss in any details the important issue of the determination of screened 
interaction parameters in the solid-state, which is still a lively topic of current research. 
On the theoretical side,  progress has been achieved using the first-principles constrained-RPA 
method~\cite{aryasetiawan_cRPA_prb_2004} and its recent developments. 
This approach has emphasized that interaction parameters (especially $U$ or $F^0$) are actually functions 
of the energy scale at which they are considered, and also depend on the set of states which are retained in the 
effective description of the solid (e.g. on the energy window used to construct appropriate Wannier functions). 
Physically, the energy-scale dependence comes from screening effects. 
At high-energy the unscreened values associated with an isolated atom are found: 
the monopole Slater integral $F^0$ (and hence $U$ and $\Up$) are of 
order $15-25\eV$. 
Screening reduces this value considerably, down to a few eV's at low-energy in the solid. 
The exchange integral, in contrast, does not involve the monopole contribution $F^0$, 
but only the two higher-order multipoles $F^2$ and $F^4$. 
Because of this, it was pointed out that the Hund's rule coupling  
is only reduced by $20-30$\% when going from the atom to the solid,  
see e.g. the pioneering work of van der Marel and Sawatzky~\cite{vandermarel_sawatzky_prb_1988,vandermarel_phd_1985}. 
While the Hartree-Fock (unscreened) value for a 3d transition element with atomic number $Z$ reads 
$J_H^{\mathrm{at}}=0.81+0.080(Z-21)\eV$, these authors estimated 
the screened  $J_H=0.59+0.075 (Z-21)\eV$ (with $C_d\simeq 0.52 J_H$ in both cases). 
This varies from $J_H\simeq 0.59\eV$ up to $J_H\simeq 1.15\eV$ as one moves along the 3d series from Sc to Ni 
(note that for a \ttg shell $J\simeq 0.77 J_H$). 
We also note, given the Hartree-Fock value $F^0_{\mathrm{at}}=15.31+1.5(Z-21)\eV$,  that 
$J_H^{\mathrm{at}}/F^0_{\mathrm{at}}\simeq 0.053$ is fairly constant along the series.
It is thus reasonable to expect that the ratio $J/U$ 
for a \ttg shell  
is also approximately constant among early transition-metal oxides~\footnote{Using for example a reduction 
of $F_0$ by screening down to $20\%$ of its atomic value, 
one obtains using the above expressions $J/U\simeq 0.13$ for a \ttg shell.}. 

\section{Energetics of the Mott gap}
\label{sec:mottgap} 

The Hund's rule coupling affects the energetics of charge transfer in a major way, in a manner 
that depends crucially on the filling of the shell. 
This effect is already visible for an isolated atom, as noted by van der Marel and Sawatzky~\cite{vandermarel_sawatzky_prb_1988,vandermarel_phd_1985}. 
It has direct consequences for the magnitude of the Mott gap in the solid-state context, as discussed below.  

Consider first an isolated shell with $N$ electrons. We are interested in the energetic cost for changing the valence of two isolated atoms 
from their nominal electron numbers to the state with $N-1,N+1$, i.e. transferring one electron from one of the atoms to the other.
This energy cost reads:
\begin{equation}
\Gapat\,=\,E_0(N+1)+E_0(N-1)-2E_0(N) = [E_0(N+1)-E_0(N)]-[E_0(N)-E_0(N-1)]
\label{eq:atomic_gap}
\end{equation} 
with $E_0$ the ground-state energy of the shell with $N$ electrons. The last expression emphasizes that this is 
the difference between the affinity and ionization  energies. 

For simplicity, we will base the discussion on the Kanamori hamiltonian (\ref{eq:ham_kanamori}) 
appropriate for example to a \ttg shell.  The ground-state energy of this hamiltonian 
can be obtained by considering simply the density-density terms. Consider the state in the (degenerate) 
ground-state multiplet with maximal $S^z$ (=$+N/2$ for $N\leq M$, = $M-N/2$ for $N\geq M$), 
consistent with Hund's rules.
For example, for 3 orbitals: $|\spinup,\spinup,0\ket$ for $N=2<M$, 
$|\spinup,\spinup,\spinup\ket$ for $N=M=3$ and $|\spinup\spindown,\spinup,\spinup\ket$ for $N=4$.
It is clear that the exchange and pair-hopping terms have no action on those states. So, for $N\leq M$, the ground-state 
energy involves only the pairwise interaction between parallel spins: $E_0(N)=(\Up-J)N(N-1)/2=(U-3J)N(N-1)/2$. As long as 
$N<M$ (less than half-filled shell), the expression of the atomic gap (\ref{eq:atomic_gap}) involves only states with 
energies of this form. Hence $U_{\rm{eff}}=U-3J$ plays the role of the effective Hubbard interaction (which is seen to 
be {\it reduced} by $J$) and the atomic gap reads:
\begin{equation}
\Gapat\,\equiv\,\Ueff\,=\,\Up-J\,=\,U-3J\,\,\,,\,\,\,(N<M\,\,\rm{or}\,\, N>M)
\label{eq:atomic_gap_non_half_filled}
\end{equation}
with the expression for $N>M$ stemming from particle-hole symmetry. In contrast, for a half-filled shell, the excited state 
with $N+1=M+1$ electrons involves one doubly-occupied orbital, and hence its energy is pushed up. 
Counting the number of each types of pairs, it reads: 
$E_0(M+1)=(\Up-J)\times M(M-1)/2 + U\times 1 + \Up\times (M-1)=(\Up-J)M(M+1)/2+(U-\Up+MJ)$. The last expression emphasizes that 
the energy of this state is increased by $U-\Up+MJ$, as compared to the value it would have if all interactions would be between parallel 
spins. Hence, the Mott gap becomes~\footnote{For a generalization including spin-orbit, see Ref.~\cite{meetei_doublepero_2012}.}: 
 \begin{equation}
 \Gapat\,\equiv\,\Ueff\,=\,
 (\Up-J)\,+\,(U-\Up+MJ)\,=\,U+(M-1)J\,\,\,,\,\,\,(N=M)
 \label{eq:atomic_gap_half_filled}
 \end{equation}
 In contrast to a generic filling $N\neq M$, the intra-atomic gap (or effective $U$) is {\it increased} 
 by Hund's coupling for a half-filled shell ($N=M$).  
 Here we have considered the Kanamori hamiltonian. Corresponding expressions for a 5-fold degenerate $d$-shell 
 with full Racah-Slater hamiltonian can be found 
 in Refs.~\cite{vandermarel_sawatzky_prb_1988,vandermarel_phd_1985,haverkort_phd_2005}, 
 with similar qualitative conclusions. 
 
\begin{figure}[!ht]
\begin{center}
\includegraphics[width=15cm]{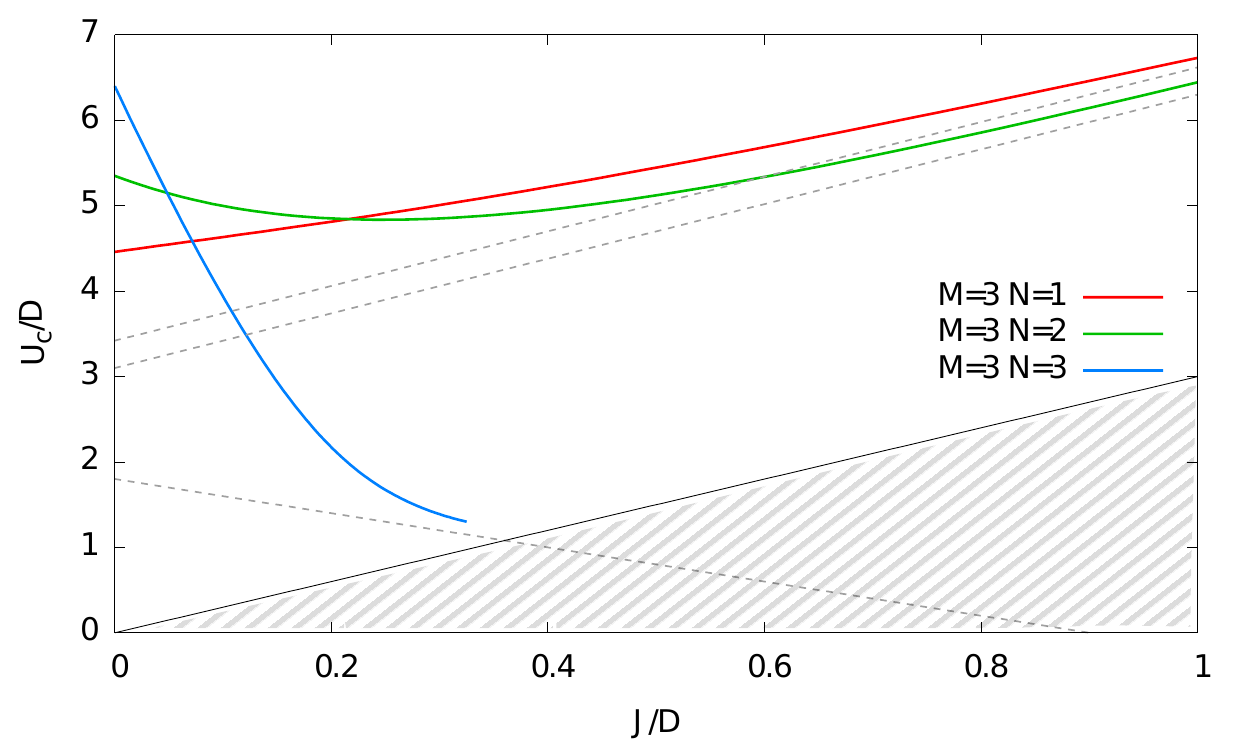}
\caption{Critical coupling separating the metallic and Mott insulating (paramagnetic) phase, as a 
function of Hund's coupling, for a Hubbard-Kanamori model of three degenerate bands 
with one (red), two (green) and three (blue) electrons per site. 
The model is solved with DMFT, with a semi-circular density of states of bandwidth $2D$ for each band. 
Dashed lines indicate the atomic-like estimates (see text). 
The shaded region corresponds to $U^\prime-J<0$ ($J>U/3$). 
See Refs.~\cite{demedici_MottHund,demedici_prl_2011}. 
\label{fig:Uc_vs_J}}
\end{center}
\end{figure}

These considerations for an isolated atom suggest that, in the solid-state context, the Hund's rule coupling 
has a strong influence on the Mott gap and on the critical coupling $U_c$ separating a metallic phase from 
a Mott insulating phase. 
Anticipating on Sec.~\ref{sec:janus}, we display on Fig.~\ref{fig:Uc_vs_J} the dependence of $U_c$ on $J$ for a 
Hubbard-Kanamori model of three degenerate bands, as obtained from DMFT calculations. 
It is seen that $U_c$ is strongly reduced as $J$ is increased, in the case of a half-filled shell ($N=3$). 
In contrast, $U_c$ is increased, with a quasi-linear dependence on $J$, for $N=1$.  
The case of two electrons (and, more generally the generic case $N\neq 1,M,2M-1$) is especially interesting, with a non-monotonous dependence: 
$U_c$ first decreases at small $J$, then increases linearly at larger~$J$. 
The strong reduction of $U_c$ by $J$ at half-filling has been discussed by 
many authors (see e.g.~\cite{han_multiorbital,
bunemann_gutzwiller_prb_1998,
koga_2orbital_Hubbard_prb_2002,
pruschke_epjb_2005,
werner07_hilowspin,nevidomskiy09}). 
%
%
The fact that $J$ enhances $U_c$ and hence makes the Mott insulating state                                       
harder to reach in the generic case of a non half-filled shell has in                                            
contrast been clearly appreciated only recently. Although implicit in the results of e.g.                                                                                              
Refs.~\cite{fresard_multiorbital_prb_1997,lombardo_Hund_prb_2005},                                               
it has been recently emphasized in Ref.~\cite{werner09} and, especially, in Ref.~\cite{demedici_MottHund}.

In order to rationalize the $J$-dependence of $U_c$ displayed on Fig.~\ref{fig:Uc_vs_J}, 
it is natural to use the atomic limit considered above and apply a criterion `\`a la Mott-Hubbard' 
for the closing of the gap~\cite{demedici_MottHund}, namely $\Ueff^{\mathrm{at}}\,=\,\Wt_{M,N}(J)$. 
In this expression, $\Wt$ is an estimate of the available kinetic energy for $N$ electrons 
hopping among $M$ degenerate orbitals. 
This leads to $U_c=\Wt_{M,N}(J)+3J$ for a non half-filled shell and $U_c=\Wt_{M,M}(J)-(M-1)J$ at half-filling.
Assuming that $\Wt$ reaches a finite value $\Wt^\infty$ at large-$J$, this yields a linear increase 
$U_c\sim \Wt_{N,M}^\infty+3J$ for $N\neq M$ and a linear decrease 
$U_c\sim \Wt_{M,M}^\infty-(M-1)J$ at half-filling. 
It is seen from Fig.~\ref{fig:Uc_vs_J} that these expressions (gray straight lines) describe the large-$J$ 
behaviour of $U_c$ quite well. 
It is also clear from Fig.~\ref{fig:Uc_vs_J} that the $J$-dependence 
of the kinetic energy $\Wt$ is crucial to account for $U_c(J)$: the extrapolations of these 
straight lines down to $J=0$ fall way below the actual value of $U_c$ at $J=0$, except in the case of a 
single electron $N=1$. 
The reason for this is that the Hund's rule coupling quenches the orbital fluctuations, which  
in turns blocks many of the hopping processes contributing to $\Wt$. 
A deeper perspective on this effect will be given in Sec.~\ref{sec:kondo} in the context of the Kondo 
problem of a magnetic impurity in a metallic host. 
This effect is particularly strong at half-filling: for $J=0$, it is well established
~\cite{lu_MIT_prb_1994,gunnarsson_orbital_1997_prb,
han_multiorbital,florens_orbital_2002_prb}
that the orbital fluctuations lead to a value of 
$U_c^{J=0}$ (and $\Wt$) which increases rapidly with orbital degeneracy $M$
\footnote{Within DMFT, the Mott-Hubbard gap-closing transition 
occurs at $U_{c1}(J=0)\propto \sqrt{M}$ and the Brinkman-Rice transition where the 
quasiparticle weight vanishes at $U_{c2}\propto M$, see Ref.~\cite{florens_orbital_2002_prb}.}. 
In contrast $\Wt$ is renormalized downwards as $J$ is turned on, and a value 
$\Wt_{M,M}\sim\Wt_{1,1}\sim W$ (with $W$ the bare bandwidth) is reached already at 
moderate values of $J$, leading to $U_c\sim W+(M-1)J$
 (L. de'Medici \& M. Capone, in preparation), as clear from Fig.~\ref{fig:Uc_vs_J}, with only
a weak dependence on  orbital degeneracy in the presence 
of a finite $J$~\footnote{Accordingly, the coexistence region $[U_{c1},U_{c2}]$ is strongly reduced by $J$~\cite{inaba_prb_2005}.}

For generic filling levels, the reduction of the kinetic energy by orbital blocking is responsible 
for the decrease of $U_c$ at small $J$, while the reduction of the atomic $\Ueff$ is responsible for the 
increase of $U_c$ at large $J$, hence the non-monotonous behaviour. 
In contrast for a single electron or hole, the orbital blocking does not apply because the 
Hund's rule coupling does not lift the degeneracy of the atomic ground-state. 

Finally, we note that at $J=0$, the largest value of $U_c$ is obtained at half-filling $N=M$ and the smallest one for a single electron 
(or hole) $N=1,2M-1$.
This is {\it reversed} at moderate and large $J$, with $U_c$ smallest for a half-filled shell 
(Fig.~\ref{fig:Zcontour}). 
Because of this effect, an insulating state is strongly favoured at half-filling. Indeed, most 
transition-metal oxides with a half-filled shell are insulators 
(e.g. \SMO, LaCrO$_3$ with three electrons in the \ttg states, see Sec.~\ref{sec:oxides_globalview}). 
The reduction of the Mott gap by Hund's coupling for a non half-filled shell was proposed early on by 
Fujimori et al.~\cite{Fujimori_V5S8} in order to explain the paramagnetic metallic character of V$_5$S$_8$ 
and its photoemission spectrum showing an exchange splitting. 

\section{Impeded Kondo screening and blocking of orbital fluctuations}
\label{sec:kondo}

We now consider a single atom hybridized with a Fermi sea of conduction electrons.  
This is the famous Kondo problem of a magnetic impurity embedded in a metallic host.  
As we shall see, the generic effect of the Hund's rule coupling is to drastically suppress the Kondo 
temperature $T_K$ below which the local moment of the atom is screened. 
This suppression is due to the combination of two effects: the blocking of 
orbital fluctuations as well as the reduction of the effective Kondo coupling 
within the low-lying multiplet selected by Hund's rule.  
For a spin-$S$ impurity, this reduction follows $J_{K,\mathrm{eff}}\propto 1/S$, 
as first derived by Schrieffer~\cite{schrieffer_japplphys_1967} (see also Blandin, Ref.~\cite{blandin_japplphys_1968}).  
The Kondo temperature being exponential in $J_K$ thus drops exponentially with $S$, as
indeed observed experimentally for metals hosting transition-metal impurities with  
different spin values~\cite{daybell_rmp_1968}.
A systematic study of the suppression of the Kondo scale by the Hund's rule 
coupling was first performed by Okada and Yosida~\cite{okada73}.

In Ref.~\cite{dworin_prl_1970}, Dworin and Narath introduced a generalization of the Anderson 
impurity model for $M$ orbitals (e.g.  $M=2l+1$ angular momentum channels) which takes into account 
Hund's rule physics in a minimal way (see also~\cite{anderson_physrev_1961,caroli_hund_prl_1969}). This reads: 
\begin{equation}
H_{\mathrm{DN}}= \sum_{\vk}\sum_{m=1}^M\sum_{\sigma=\spinup\spindown}
\left(\ek c^{\dagger}_{\vk m \sigma}c_{\vk\sigma} +  V_{\vk m} c^\dagger_{\vk m\sigma} d_{m \sigma} + 
V_{\vk m}^* d^\dagger_{m\sigma}c_{\vk m\sigma}\right) \,+\, 
H_{\rm{at}}
\label{eq:ham_DN}
\end{equation}
with the atomic term: 
\begin{eqnarray}
 H_{\mathrm{at}}
 &=&\frac{U-J}{2}\sum_{m_1 m_2 s_1 s_2}d^\dagger_{m_1
    s_1} d^\dagger_{m_2 s_2} d_{m_2 s_2} d_{m_1 s_1} + 
\frac{J}{2}  \sum_{m_1 m_2 s_1 s_2} d^\dagger_{m_1 s_1} d^\dagger_{m_2 s_2}d_{m_1 s_2} d_{m_2 s_1} \nonumber \\ 
  &=&  (U-\frac{3}{2}J)\frac {\hN_d(\hN_d-1)}{2}-J \vec{S}_d^2 +\frac{1}{4}J \hN_d 
  \label{eq:ham_DN_at}.
\end{eqnarray}
where $\hat{N}_d$ and $\vec{S}_d$ are, respectively, the total charge and spin operators of the $d$-shell as above.  
The atomic part of the Dworin-Narath hamiltonian is rotationally invariant and 
coincides with the generalized Kanamori hamiltonian with appropriately chosen parameters, 
as discussed in Sec.\ref{sec:hund} and Appendix~\ref{sec:appendix} (Eq.~(\ref{eq:ham_dworin_2orbital})). 

For $J=0$ the model has full $SU(2M)$ symmetry. Coqblin and Schrieffer (CS) pioneered~\cite{coqblin_schrieffer_1969} the study of impurity 
models with enhanced orbital symmetry by considering the hamiltonian:
\begin{equation}
H_{\mathrm{CS}} = \sum_{\vk\alpha} \ek
c^\dagger_{\vk\alpha}c_{\vk\alpha}\,+\, 
J_K\,\sum_{\vk\vk^\prime}\sum_{\alpha\beta}
c^\dagger_{\vk\alpha}c_{\vk^\prime\beta} \hat{S}_{\beta\alpha}
\label{eq:ham_coqblin}
\end{equation}
%
%
where $\alpha=\{m,\sigma\}$ and $\beta$ are $SU(2M)$ indices and $\hat{S}_{\alpha\beta}$ is the impurity operator corresponding to 
a specific irreducible representation of $SU(2M)$.  At large $U$ and $J=0$, a Schrieffer-Wolff transformation maps the 
Dworin-Narath hamiltonian (\ref{eq:ham_DN},\ref{eq:ham_DN_at}) onto the CS hamiltonian when the number of electrons 
$N_d=\sum_{m\sigma}d^\dagger_{m\sigma}d_{m\sigma}$ is constrained to be an integer 
(then $\hat{S}_{\alpha\beta} \propto d^\dagger_\alpha d_\beta$). 

A well established result for the CS model~\cite{hewson_book_1993} is
that the Kondo temperature is {\it enhanced} by the orbital
degeneracy. For a half-filled shell $N_d=M$, one has
\begin{equation}
T_{K,M}^{J=0}/D\,=\exp(-1/2M\rho J_K)\,=\,\left(T_{K,1}/D\right)^{1/M}
\label{eq:TK_CS} 
\end{equation}
and a similar enhancement applies for all values of $N_d$ 
(see Ref.\cite{nishikawa_hewson_prb_2010_2} for a detailed 
comparison of $N_d=1$ and $N_d=2$ in the case of $M=2$ orbitals).  
In this expression, $\rho$ is the conduction electron density 
of states (per orbital and spin), $D$ is a high-energy cutoff (e.g. the bandwidth for the CS model, or $\sim\sqrt{U\Gamma}$ for 
the Anderson model) and $T_{K,1}$ is the usual Kondo temperature for a single orbital 
$T_{K,1}/D \sim \exp(-1/2\rho J_K)$. 
Intuitively, the enhancement of $T_K$ for $J=0$ occurs because the
conduction electrons can exchange both spin and orbital momentum with
the impurity spins, which enhances the corresponding gain in exchange energy. 
The enhancement of the Kondo temperature due to 
orbital degrees of freedom has been investigated intensively in
mesoscopic systems. $SU(4)$ symmetry with entangled spin and orbital degree
of freedom in carbon nanotubes and quantum dots has been discussed
theoretically \cite{borda_prl_2003,lehur_prb_2003,zarand_ssc_prb_2003,
  galpin_krishnamurthy_prl_2005, mravlje_prb_2006} and its effects
have been observed experimentally~\cite{jarillo-herrero_nature_2005}.

\begin{figure}[!ht]
\begin{center}
\includegraphics[width=10cm]{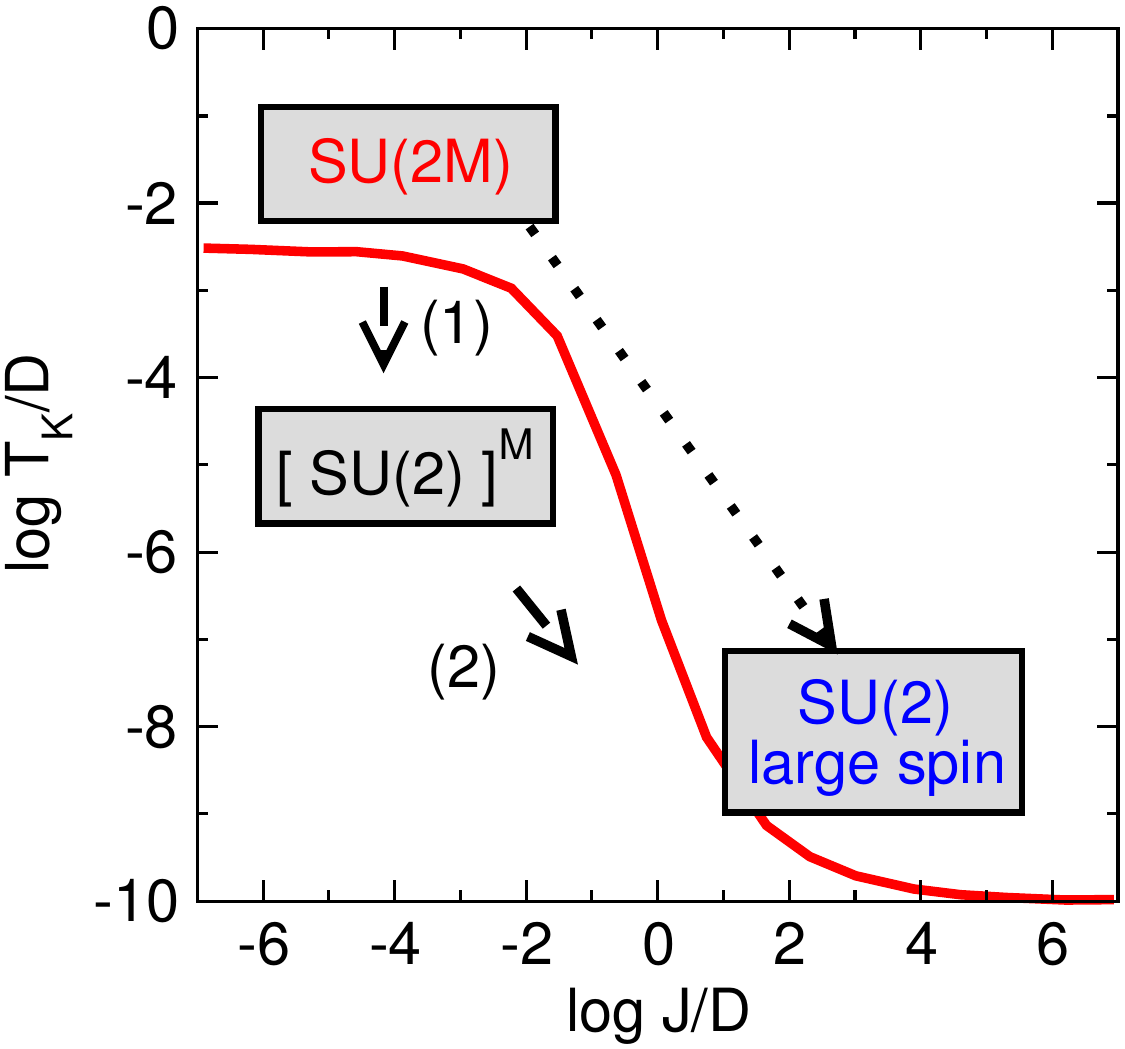}
\caption{Kondo temperature $T_K$ as a function of Hund's
  coupling $J$ for the 2-orbital `Coqblin-Schrieffer-Hund' model, plotted on a log-log scale.
  The data (red line) are from the poor-man's scaling
  analysis in Ref.~\cite{yanase97}. $J$ suppresses the
  Kondo temperature and lowers the symmetry of the problem (see text).
\label{fig:suppression_Tk_schematic}}
\end{center}
\end{figure}

A non-zero $J$ breaks the $SU(2M)$ symmetry down to $SU(2)_S\otimes SU(M)_O$. 
It drastically modifies the physics and reduces $T_K$, as first discussed systematically
by Okada and Yosida~\cite{okada73} for model (\ref{eq:ham_DN},\ref{eq:ham_DN_at}) in the large-$U$ limit.
These authors performed a Schrieffer-Wolff transformation for a fixed
integer $N_d$ and obtained a `Coqblin-Schrieffer-Hund' model which 
basically consists in adding a term $-J\vec{S}^2$ to
(\ref{eq:ham_coqblin}). 
This model was then analyzed by further taking $J\to 0$ or $J\to \infty $ and projecting onto an appropriate subspace,
which depends on $N_d$.  A variational wave-function approach was used
and the resulting binding energies were related to the Kondo
temperature.  For a half-filled shell $N_d=M$ and in the limit of
large $J$ the Kondo temperature is strongly reduced: 

\begin{equation}
T_{K,M}/D=\,\exp(-M/2\rho J_K)\,=
\left(T_{K,M}^{J=0}/D\right)^{M^2}=\left(T_{K,1}/D\right)^M 
\label{eq:TK_largeJ}
\end{equation}
A similar reduction was found also for $N_d=M\pm 1$ ($M>2$)~\cite{okada73}, and in this case
it was furthermore observed that the orbital fluctuations are quenched
at a larger energy scale than the spin fluctuations.  
The narrowing of the Kondo resonance and suppression of $T_K$ due to $J$
was also studied using NRG for the two-orbital model 
by Pruschke and Bulla~\cite{pruschke_epjb_2005}.
$J$ thus takes the system away from the point of high symmetry and high $T_K$, see 
Fig.~\ref{fig:suppression_Tk_schematic}.

Again, the reduction of the Kondo scale can be understood intuitively. For a half-filled shell, 
the (degenerate) atomic ground-state for $J\neq 0$ has a large spin $S=M/2$ and a vanishing 
angular momentum $L=0$ (see Table~\ref{table:spectrum_t2g} for 3 orbitals). Hence, all 
the orbital exchange energy applying to the $J=0$ case is lost here because orbital exchange 
processes are blocked. Furthermore, at large $J$, 
spin-exchange processes are restricted to the ground-state subspace with $S=M/2$ 
(with therefore a smaller degeneracy than the $J=0$ ground-state subspace). 
The impurity spins in each orbital channel acts in this subspace as $\vec{S}_m\sim \vec{S}/M$, the 
proportionality factor $1/M$ being a Clebsch-Gordon coefficient. 
(For a related analysis in the case of an antiferromagnetic Hund's coupling, see 
Ref.~\cite{zitko_jphys_2010}). 
Hence, the effective Kondo coupling 
$J_{K,\mathrm{eff}}=J_K/M$ is reduced, as first observed by Schrieffer~\cite{schrieffer_japplphys_1967} 
(see also \cite{blandin_japplphys_1968}). 

This is quite transparently seen, following 
Refs.~\cite{jayaprakash81,jones87,kusunose97,yotsuhashi01,nevidomskiy09}, by considering a `composite spin' Kondo (CSK) 
hamiltonian as a starting point:
\begin{equation}
H_{\mathrm{CSK}}\,=\,
\sum_{\vk m\sigma}\ek c_{\vk m\sigma}^\dagger c_{\vk m \sigma}
+J_K\sum_m \vec{S}_m \cdot \vec{\sigma}^{\,c}_m 
- J \left( \sum_m\vec{S}_m\right)^2
\label{eq:ham_kusu}
\end{equation}
describing $M$ spin-1/2 impurities $\vec{S}_m$,  
each Kondo-coupled to the spin density $\vec{\sigma}^{\,c}_m=\sum_{\vk \alpha\beta} c_{\vk m\alpha}^\dagger \vec{\sigma}_{\alpha \beta} c_{\vk m \beta}$ of an independent
bath (with $\vec{\sigma}$ the Pauli matrices). The spins are coupled by the Hund term favoring $S=M/2$. 
A crucial difference with the Dworin-Narath (\ref{eq:ham_DN},\ref{eq:ham_DN_at}) and Coqblin-Schrieffer-Hund hamiltonians is that 
there are no orbital degrees of freedom here. The Kondo coupling is diagonal 
$\sim J_K\,\delta_{m\mp}$, and the $J=0$
hamiltonian thus has a smaller $[\mathrm{SU(2)}]^M$ symmetry. 
When $J$ becomes larger than the other scales in the problem, 
the large spin is formed. Within this subspace $\vec{S}_m\sim\vec{S}/M$,  
and the Hamiltonian (\ref{eq:ham_kusu}) is equivalent to the 
$M$-channel Kondo problem with spin $S=M/2$ and 
$J_{K,\mathrm{eff}}=J_K/M$. 
The low-energy fixed point is a Fermi-liquid with an exactly screened impurity spin since $M=2S$. 
\begin{figure}[!ht]
\begin{center}
\includegraphics[width=15cm]{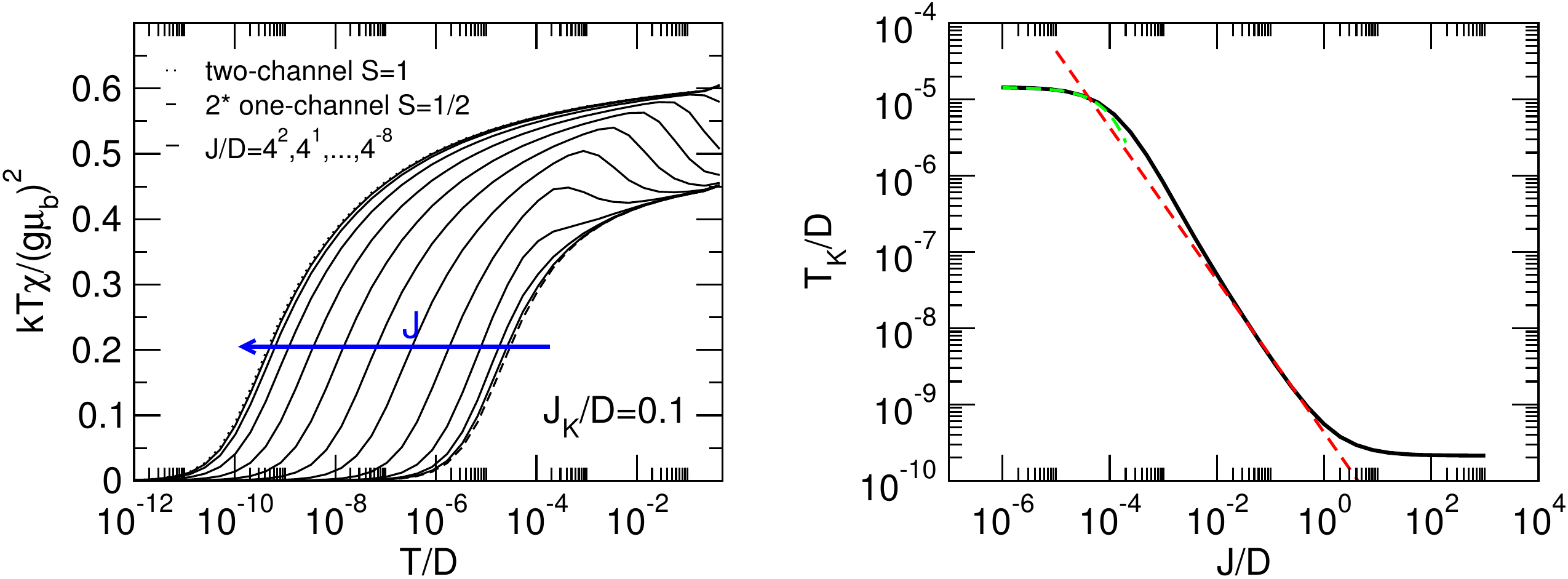}
\end{center}
\caption{NRG results for the composite-spin Kondo 
hamiltonian (\ref{eq:ham_kusu}) with $M=2$. Data
  from Ref.~\cite{demedici_prl_2011}.  Left panel: the impurity
  contribution to the magnetic susceptibility for several values of
  the Hund's rule coupling~$J$.  The behaviour evolves from that of two
  independent Kondo problems to that of the $S=1$ Kondo problem as $J$
  is increased.  Note the increase in $\chi$ at intermediate values of $J$ 
  (cf. Fig.~\ref{fig:Nevidomskiy}).  
  Right panel: the corresponding Kondo temperature as
  a function of $J$. At small $J$ an exponential dependence
  $T_K=T_{K,1}\exp(-J/8.4T_{K,1})$ with $T_{K,1}=T_K(J=0)$ (green dotted) and at larger $J$ a
  power-law $T_K \propto 1/J$ (red dashed line) are found.
\label{fig:chi}}
\end{figure}
A numerical renormalization-group (NRG) study \cite{demedici_prl_2011} (Fig.~\ref{fig:chi})
of the CSK hamiltonian (\ref{eq:ham_kusu}) shows that the Kondo temperature 
of this model is indeed reduced according to (\ref{eq:TK_largeJ}) in the large-$J$ limit.

We finally discuss the behaviour at intermediate values of  Hund's coupling, which is of direct 
interest in view of applications to the transition-metal oxides discussed later in this paper, in which typically 
$J\sim U/6 < D$. 

\begin{figure}[!ht]
\begin{center}
\includegraphics[width=12cm]{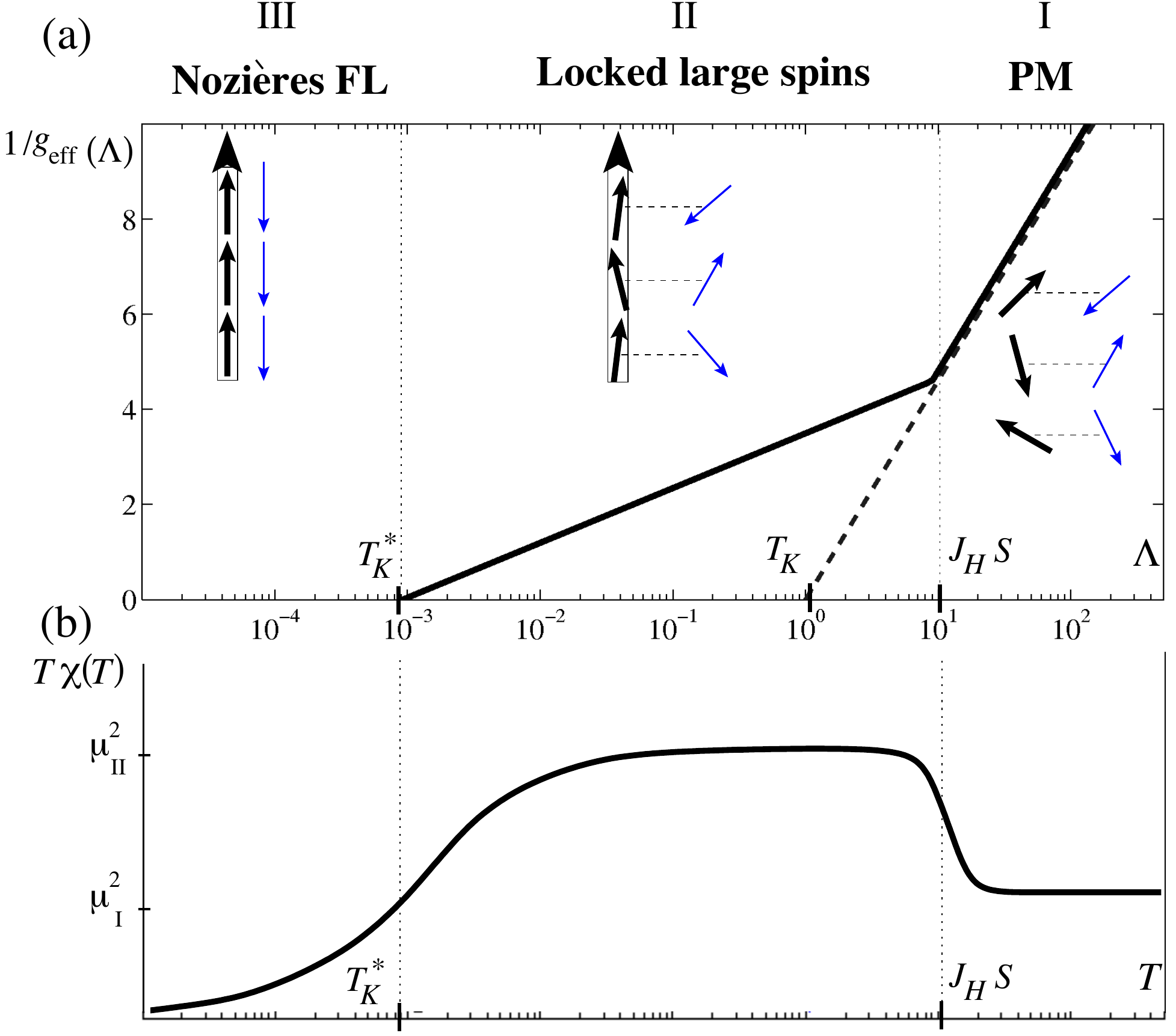}
\end{center}
\caption{Composite-spin Kondo model (Eq.~\ref{eq:ham_kusu}). 
(a) Schematic behavior of the running coupling constant 
  $g_\mathrm{eff}=J_K(\Lambda) \rho M_\mathrm{eff}$ with
  $M_\mathrm{eff}=1$ in region I and $g_\mathrm{eff}=M$ in regions II
  and III. The boundary between I and II is at the scale of Hund's
  coupling. The Kondo temperature is reduced due to the slower scaling in
  region II. (b) Schematic dependence of the effective moment. The large moment formed in
  region II is screened at a reduced temperature scale.
  Reproduced from Ref.~\cite{nevidomskiy09}. 
\label{fig:Nevidomskiy}} 
\end{figure}

In Refs.~\cite{jayaprakash81,nevidomskiy09} this was analyzed for the CSK hamiltonian using 
perturbative RG, leading to 
an explicit expression for $T_K(J)$. 
The RG flow was separated into two regions (see the 
schematic plot in Fig.~\ref{fig:Nevidomskiy}). At high energies (I)  
$\Lambda > J$ and the impurity spins are not yet locked into the large-spin state. 
There, $J_K(\Lambda)$ grows with diminishing $\Lambda$ as in the single-channel single-impurity case. 
In region (II) $\Lambda <J$, the large spin is assumed to be established. The
key point is that in this region the Kondo coupling is reduced by a
factor $1/M$ and the speed at which it flows is reduced by the same
factor.
This can be summarized in a single scaling equation (to two-loop order): 
$d\geff/d\ln\Lambda=-2\geff^2/M+2\geff^3/M$ 
for the effective coupling constant 
$\geff=\rho J_K(\Lambda)M_{\mathrm{eff}}$ with the effective number of channels 
$M_{\mathrm{eff}} = 1$ in region (I) and $M_{\mathrm{eff}}=M$ in region (II). 
Because of the slower scaling in region (II), the screened Kondo
regime (III), signalled in a perturbative RG treatment by a diverging
coupling constant, occurs at a scale much smaller than the
single-impurity scale $T_{K,1}$:
\begin{equation}
T_{K,M} =T_{K,1} \left(\frac{T_{K,1}}{J S} \right)^{M-1}\,\,\,\,\,\,(\mathrm{for}\,\,H_{\mathrm{CSK}}) 
\label{eq:nevicoleman_TK}.
\end{equation}
with $S=M/2$. This is Eq.~(\ref{eq:TK_largeJ}), but with $J S$ playing the role of the
high-energy cutoff $D$.  
This RG analysis emphasizes that starting with a small enough $J$, the screening process 
in the CSK 
model proceeds first by the formation of a large spin $S$, which is then 
eventually screened at a lower scale (Fig.~\ref{fig:Nevidomskiy}). NRG studies for 
$M=2$ displayed in Fig.~\ref{fig:chi} confirm this expectation. 
One observes there an initial exponential reduction of $T_K$ at small $J$, followed by a 
power-law dependence which at quite large $J$ does match $\sim 1/J^{M-1}=1/J$.

When considering the original model  (\ref{eq:ham_DN},\ref{eq:ham_DN_at}) or the Coqblin-Schrieffer-Hund model, it is qualitatively appealing to 
think of the reduction of $T_K$ as following a two-stage process (Fig.~\ref{fig:suppression_Tk_schematic}): 
first a projection onto a subspace described by the 
CSK hamiltonian in which the different orbital channels are decoupled, 
followed by a second stage in which a large spin is formed and eventually screened at a low energy scale. 
However, it is not guaranteed that this two-stage process does apply in general, and a direct route may apply instead 
(dashed arrow on Fig.~\ref{fig:suppression_Tk_schematic}).  
Indeed, in the original model, at large scales $\Lambda>J$, the RG flow
goes as in the $SU(2M)$ symmetric model.  For $\Lambda\lesssim J$, the quenching of the orbital
fluctuations and the emergence of the high-spin state occur \emph{simultaneously}. 
There is no energy scale at which the system is represented
by $M$ independent spins undergoing single-channel Kondo scaling.
As a result, expression (\ref{eq:nevicoleman_TK}) for the reduction of $T_K$ at intermediate $J$ 
for the CSK 
model cannot be trusted in general for the original model (\ref{eq:ham_DN},\ref{eq:ham_DN_at}) or the Coqblin-Schrieffer-Hund model. 
Indeed, the poor man's scaling study of Ref.\cite{yanase97} for $M=2$ reproduced 
in Fig.~\ref{fig:suppression_Tk_schematic} suggests a $1/J^2$ dependence, instead of $1/J$  as in (\ref{eq:nevicoleman_TK})  
while NRG studies by one of us (J.M., unpublished) yield an even stronger power-law. 
NRG studies of the Dworin-Narath model (\ref{eq:ham_DN},\ref{eq:ham_DN_at}) were also performed in
Refs.~\cite{nishikawa_hewson_prb_2010_1,nishikawa_hewson_prb_2010_2}
and an exponential dependence of $T_K$ on $J$ was reported. We expect
that this is because rather small values of $J$ were explored there,
and that an initial exponential suppression followed by a power-law at
larger $J$ is the generic behaviour, as shown on
Fig.~\ref{fig:suppression_Tk_schematic}. Explanation of the
exponential regime and the characterization of precise power-laws (in
particular in the case where the impurity has nonvanishing orbital
momentum) remains to be worked out.

\section{Dynamical Mean-Field Theory: solids viewed as embedded atoms}
\label{sec:dmft}

Having considered isolated atoms (Sec.~\ref{sec:mottgap}), and a single impurity atom in a host metal (Sec.~\ref{sec:kondo}), 
we now move to a full solid - a periodic array of atoms exchanging electrons.
The main message of this article is that the intra-atomic correlations associated with Hund's coupling 
play a crucial role also in this context. 
Dynamical Mean-Field Theory (DMFT) is currently the most appropriate framework in which these effects can 
be revealed and studied~\cite{georges_review_dmft,kotliar_dmft_physicstoday}. 
Indeed, while more traditional approaches view a solid as an inhomogeneous electron gas to which 
interactions are later added, DMFT gives central importance to the fact that, after
all, solids are made of atoms and that an atom is a small
many-body problem in itself with e.g. a multiplet structure which must be properly taken into account.

\begin{figure}[!ht]
\begin{center}
\includegraphics[width=12cm]{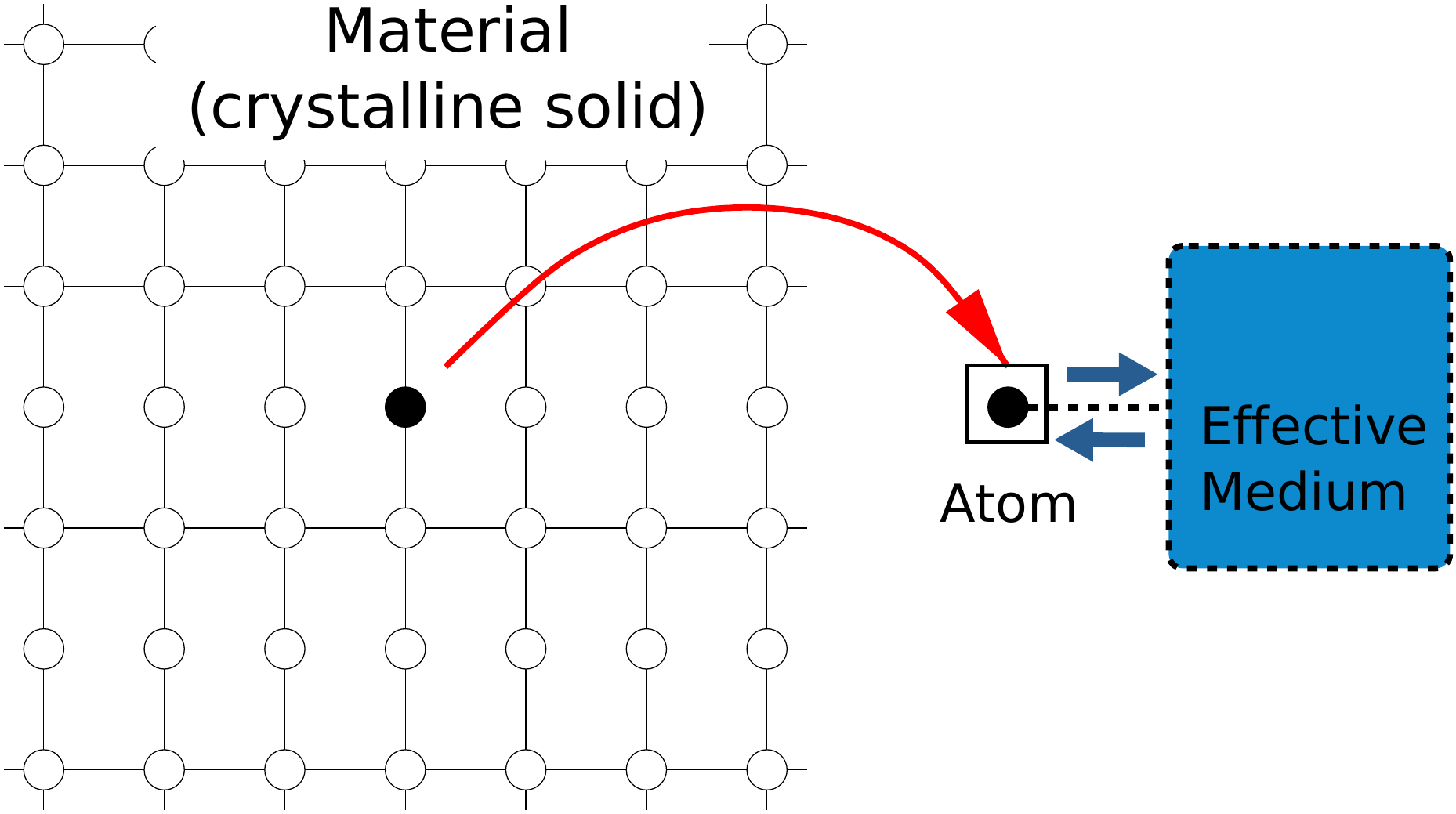}
\caption{
The Dynamical Mean-Field Theory (DMFT) concept. 
A solid is viewed as an array of atoms exchanging electrons, rather than as a gas of 
interacting electrons moving in an inhomogeneous potential.  DMFT replaces the solid by a single atom exchanging 
electrons with a self-consistent medium and takes into account many-body correlations on each atomic site. 
\label{fig:dmft}
}
\end{center}
\end{figure}

DMFT describes the transfer of electrons between atoms in the solid by focusing on a single atomic site, and by 
viewing the atom on this site as hybridized with an effective medium with which these electronic transfers 
take place (Fig.~\ref{fig:dmft}). 
This effective medium must obviously be self-consistently related to the rest of the solid. 
In more technical terms, the main physical observable on which DMFT focuses is the single-electron Green's function 
$G_{m\mp}(\omega)$ (or spectral function $A=-\mathrm{Im}G/\pi$) for a given atomic shell, e.g. the d-shell of an oxide. 
This observable is represented as that of an atomic shell coupled to the effective medium via a 
hybridization function $\Delta_{m\mp}(\omega)$ which can be viewed as an energy-dependent (dynamical)  
generalization of the Weiss effective-field concept to quantum many-body systems. 
The key assumption is that the self-energy $\Sigma_{m\mp}(\omega)$ of this effective quantum impurity model can 
be used as a (local) approximation of the full self-energy of the solid. 
The Dyson equation, projected onto the local orbitals $\chi_m(\vr)$ defining the correlated shell, then 
yields a self-consistency condition for $G_{m\mp}$, which also determines $\Sigma_{m\mp}$ and $\Delta_{m\mp}$.
%
%
DMFT has been successfully combined with DFT-based electronic structure calculations for real materials, in which 
case the all-electron charge density $\rho(\vr)$ is the other key observable together with $G_{m\mp}(\omega)$ 
(for reviews, see e.g. Refs~\cite{kotliar_review_rmp_2006,held_review_advphys_2007,georges_strong,
imada_review_jpsj_2010}). 

The energy-dependence of the dynamical mean-field $\Delta_{m\mp}(\omega)$ is of central physical 
importance. 
Indeed, in strongly-correlated materials, electrons are `hesitant' entities with a dual character. 
At high-energy they behave as localized. At low-energy in metallic compounds they 
eventually form itinerant quasiparticles, albeit with a strongly suppressed spectral weight. 
DMFT describes the high-energy behaviour by taking full account of the multiplet structure of the atomic shell, and of its broadening 
by the solid-state environment. The latter is encoded in the high-frequency behaviour of $\Delta_{m\mp}(\omega)$, which 
gives for example their widths to Hubbard satellites. 
At low-energy, the key issue is whether the degeneracy of the ground-state multiplet is fully lifted by 
the solid-state environment. 
In metallic systems, the effective hybridization $\mathrm{Im}\Delta_{m\mp}(\omega)$ does not vanish at low-energy 
(in contrast to a Mott insulator, where it displays a gap). 
As a result, Kondo screening of the ground-state multiplet can take place. 
In this context, this self-consistent Kondo screening is the local description of electron transfer processes which screen out the multiplet 
structure in the metallic ground-state. 
For example in the simplest context of a single orbital Hubbard model, a twofold degenerate spin-$1/2$ 
local moment is found in the parmagnetic Mott insulating phase, while it is Kondo-screened into a singlet 
in the metallic phase. 

In most cases, this results in a description of the low-energy
excitations in the metallic phase in terms of quasiparticles of a
local Fermi-liquid. These are characterized by three key quantities:
their quasiparticle weight $Z$, effective mass $m^*$ (or renormalized
Fermi velocity $v_F^*/v_F^{\mathrm{band}}$) and lifetime
$\hbar/\Gamma_{\mathrm{qp}}$.
For a single-orbital model those are given by: 
$Z^{-1}=m^*/m_{\mathrm{band}}=v_F^{\mathrm{band}}/v_F^*=1-\partial\Sigma/\partial\omega|_{\omega=0}$ and 
$\Gamma_{\mathrm{qp}}=Z|\mathrm{Im}\Sigma(\omega=0)|$, with proper matrix generalization to the 
multi-orbital context. 
Note that, in a local Fermi liquid with a momentum-independent self-energy, the effective mass  enhancement coincides with $Z$ (in a multi-orbital context however, the 
renormalization of the Fermi velocity can depend on the point along the Fermi-surface through the momentum-dependence of 
the orbital character of the band).  

Fermi-liquid behaviour applies below a scale $\TFL$ which is related to the self-consistent Kondo scale (although 
significantly smaller quantitatively). 
%
According to DMFT, the study of multi-orbital Kondo impurity models in Sec.~\ref{sec:kondo} is thus directly relevant to 
a full periodic solid. 
The results established in Sec.~\ref{sec:kondo} for the Kondo temperature of an impurity coupled to a structureless bath cannot be 
directly applied to DMFT studies of a correlated solid however.  Indeed, the energy-dependent structure of the self-consistent 
hybridization must be properly taken into account (in more technical terms, one has to deal with an intermediate-coupling 
Kondo problem). 
Nonetheless, the strong suppression of the Kondo scale by Hund's coupling implies that low values of the 
Fermi liquid 
scale will be observed in the solid-state context, as detailed below in Sec.~\ref{sec:janus}.
The Landau description of quasiparticles is thus fragile in a strongly correlated metal. 
Quasiparticle excitations may survive in a range of temperature above $\TFL$, but their 
lifetime no longer obeys the $T^{-2}$ law of Fermi liquid theory.  
Because $\TFL$ is low, the understanding of the 
metallic state for $T>\TFL$ is often of direct experimental relevance. 
At very high temperature, the effective DMFT hybridization $\Delta(\omega,T)$ 
is small and the physics of independent atoms is recovered, while at low temperature $\Delta(\omega\simeq 0,T\ll\TFL)$ saturates 
and local Fermi liquid coherence is established. In between, quasiparticles become gradually less coherent and strong transfers 
of spectral weight are observed.
By bridging the gap between isolated atoms and the low-energy coherent regime, DMFT is currently the tool of choice to handle 
the entire crossover from a Fermi liquid at low temperature all the way to a bad metal incoherent regime at high temperature. 
This is, in our view, essential to the physical understanding of  many strongly correlated materials.

\section{Hund's correlated materials and the Janus-faced influence of the Hund's rule coupling}
\label{sec:janus}

In this section, we expose the main physical point of this article: 
the Hund's rule coupling 
has generically a conflicting effect on the physics of the solid-state. On the one hand it increases the
critical $U$ above which a Mott insulator is formed (Sec.~\ref{sec:mottgap}), on the other hand
(Sec.~\ref{sec:kondo}) it reduces the temperature and the energy
scale below which a Fermi-liquid is formed, leading to a (bad) metallic regime 
in which quasiparticle coherence is suppressed.

This occurs for any occupancy, the two exceptions
being a half-filled shell or a shell with a single electron or a single hole. 
In the former case, Hund's coupling strongly decreases the Mott critical coupling 
and suppresses the coherence scale, both effects leading to a more  correlated behaviour. 
In the latter case, Hund's coupling tends to decrease correlation effects by enhancing
$U_c$ without a strong effect on the coherence scale 
since the ground-state degeneracy of the isolated atom is unchanged by J in this case 
(Table~\ref{table:spectrum_t2g}).  
In all other cases the Hund's rule coupling  has two faces, like the Roman god Janus. 
%
This implies that a large class of materials 
display hallmarks of strong electronic correlations while not being close to a Mott insulating state. 

\subsection{Simplest model: three degenerate orbitals}
\label{sec:3orbital_model}

The simplest model in which 
the Janus behavior occurs is the Hubbard-Kanamori model of three degenerate bands 
described by the hamiltonian:
\begin{equation}
H\,=\,-\sum_{ij,m\sigma} t_{ij}\, d^\dagger_{im\sigma}d_{jm\sigma}\,+\,\sum_i H_{K}(i)
\end{equation}
with  $H_K=(U-3J)\hn_i(\hn_i-1)/2-2J\vec{S}_i^2-J/2\vec{L}_i^2$ the rotationally invariant 3-orbital interaction 
of Eq.~(\ref{eq:ham_t2g_NSL}). 
This describes, for example, transition-metal oxides with cubic symmetry
and a partially filled \ttg shell well separated from the empty \eg shell. 
This hamiltonian has been studied using DMFT by several authors, e.g.~\cite{werner08,werner09, demedici_prl_2011}.

In Fig.~\ref{fig:Zcontour},  we display as a colour map the value of the
quasiparticle spectral weight $Z$ as a function of the filling of the
shell ($n=\langle\hn_i\rangle$) and of the strength of the coupling
$U/D$, with $D$ the half-bandwidth.  
We have used a fixed ratio $J/U=0.15$ (cf. discussion at the
end of Sec.~\ref{sec:mottgap}) and a semi-realistic $t_{2g}$ density-of-states 
(inset of Fig.~\ref{fig:Zcontour}). 
The Mott insulator is indicated by thick vertical bars. 
Long-range ordering was suppressed in these calculations:  
Fig.~\ref{fig:Zcontour} 
displays properties of the paramagnetic state only and 
is {\it not} a full phase diagram.

Fig.~\ref{fig:Zcontour} reveals the following interesting features. 
(i) The Mott insulating state is most stable at half-filling $n=3$ where the Mott critical coupling 
$U_c$ is at least twice smaller than at other filling levels. In contrast, $U_c$ is enhanced by 
$J$ for the other filling levels, as indicated by the arrows on Fig.~\ref{fig:Zcontour}.  
For vanishing $J$, $U_c$ would instead be largest at half-filling  (crosses on Fig.~\ref{fig:Zcontour}). 
(ii) At $n=1$ and $n=5$ correlation effects are weak except in direct proximity to the Mott state, 
i.e close to $U_c$. 
(iii) In contrast, at the `Janus' filling levels $n=2$ and $n=4$, the
white region of small $Z$ extends to quite small $U$, as pointed out 
in~\cite{demedici_prl_2011}. Strongly correlated metallic phases are thus found 
in a wide range of coupling, without direct proximity to the Mott insulating state. 
(iv) A pronounced particle-hole asymmetry is observed, with stronger correlations 
on the right-hand side of Fig.~\ref{fig:Zcontour} (larger $n$'s). 
This is due to the higher value of the \ttg density-of-states close to the Fermi level, in relation to 
the van Hove singularity. This implies smaller kinetic energy, and hence slower quasiparticles which are 
easier to localize~\cite{mravlje11}. 

It should be noted that these features are in very good agreement with the map of 
transition-metal oxides (TMOs) put forward in the pioneering work of  A.~Fujimori~\cite{fujimori_diagram_1992} 
on the basis of experimental and empirical considerations (see also \cite{imada_mit_review}). 
The calculations leading to Fig.~\ref{fig:Zcontour} taking into account the 
key physical role of  Hund's coupling provide strong 
theoretical support to  such classifications of TMOs, as discussed in details below.  

\subsection{A global view on early 3d and 4d transition-metal oxides}
\label{sec:oxides_globalview}

We now make contact with real materials and show that the physical effects revealed 
at the model level above  allow one to build a global picture of how the strength 
of electronic correlations evolves in TMOs as one moves along the $3d$ and $4d$ series. 
On Fig.~\ref{fig:Zcontour} the correlation strength of several early TMOs 
is indicated. 
For most of the metallic compounds, the specific-heat and its
enhancement over the LDA value $\gamma/\gamma_\mathrm{LDA}$ is
reliably known from experiments.  
These are positioned on Fig.~\ref{fig:Zcontour} by demanding that the value 
of $Z^{-1}$ obtained in the model calculation at the DMFT level (where $m^*/m=Z^{-1}$)  
coincides with $\gamma/\gamma_\mathrm{LDA}$.  
Materials in the same series are positioned with a slight increase of $U/D$ along the series, 
because the bandwidth diminishes and the screened value of $U$ increases slightly as the atomic number 
and hence $n$ increases\footnote{Bandwidths for cubic 3d
  TMOs are $2.6,2.5,2.4$\,eV for SrVO$_3$, SrCrO$_3$, SrMnO$_3$,
  respectively. For 113 4d series (in cubic structure), the values
  are: $3.8,3.7,3.6$\,eV for Mo, Tc, and Ru-compound,
  respectively. For 214 (in tetragonal structure), the xy (xz)
  band-widths are $3.8 (2.2), 3.6 (1.8), 3.4 (1.5),3.1 (1.3)$\,eV, for
  Mo-, Tc-,Ru-, and Rh-compound, respectively. 
}.  
As expected, significantly larger values of interactions pertain to 3d oxides.
Apart from this, only a moderate variation of $U/D$ values is needed
to account for systematics of the early TMOs~\footnote{These materials were also explicitly simulated within LDA+DMFT. 
The main experimental properties are properly reproduced with only a mild variation of the interaction parameters
~\cite{mravlje11,demedici_prl_2011,mravlje_SrTcO3_prl_2011}.} 
  
Consider first \ttg oxides of the $3d$ series \SVO, \SCO and \SMO.  These three materials share 
a similar typical coupling $U/D\simeq 3-4$ ($U\simeq 3-4\eV$, $D\simeq 1-1.5\eV$ in \ttg description). 
 Nevertheless, they have very different
physical properties.
The origin of these differences is to be found in the different nominal filling of the 
\ttg shell, by one, two and three electrons respectively. 
For materials with a half-filled $t_{2g}^3$ shell such as SrMnO$_3$ or LaCrO$_3$, 
the ratio $U/D\simeq 4$ exceeds substantially the Mott insulating
critical value for this case (which is strongly reduced by the effect of $J$). 
This explains why no metallic $3t_{2g}^3$ oxides are
known~\cite{imada_mit_review, Torrance_why}.  
In contrast, the $3t_{2g}^1$ cubic SrVO$_3$ is a moderately correlated metal with 
$\gamma/\gamma_\mathrm{LDA} \sim m^*/m\simeq 2$~\cite{imada_mit_review}.  
In this case, $U_c$ is increased by $J$ and indeed LDA+DMFT calculations explicitely 
demonstrate~\cite{demedici_prl_2011}, see also Sec.~\ref{sec:ruthenates}, 
that SrVO$_3$ would be significantly more correlated~\cite{werner09} if $J$ was 0.
For $3t_{2g}^2$ materials, within the same range of $U/D$
strongly correlated behavior caused by the Janus-faced
action of $J$ is expected.  Cr-perovskites are situated there, but
unfortunately their synthesis necessitates high pressures, which limits
the purity of the samples. The experimental data so far is
controversial: whereas initially SrCrO$_3$ was reported to be a paramagnetic
metal~\cite{chamberland67}, a more recent study finds a semiconducting
resistivity and strong dependence of magnetic susceptibility on
temperature~\cite{zhou06}.  
Overall, the series \SVO - \SCO -\SMO beautifully illustrates the importance of the Hund's coupling, and 
of the band filling as a key control parameter. 

Oxides of $4d$ transition metals are characterized by smaller values of 
$U/D\simeq 1-2$,  due to the larger bandwidths and smaller screened interaction 
associated with the more extended 4d orbitals. 
We consider the series SrMO$_3$ and Sr$_2$MO$_4$ with 
M = {Mo, Tc, Ru, Rh} (Fig.\ref{fig:Zcontour}). 
The Technetium compounds are special among those.  
Because they have a half-filled \ttg shell and given the relevant value of $U/D$, 
these materials are located very close to the insulator to metal transition.
We are not aware of transport measurements on these compounds, but a recent
study~\cite{rodriguez11} reports antiferromagnetism with a very
large N\'eel temperature $T_N\simeq 1000$~K for SrTcO$_3$. 
Indeed, model considerations suggest that the proximity to the 
Mott critical coupling leads to largest values of the N\'eel temperature. This 
qualitative observation, together with quantitative LDA+DMFT calculations supporting it, was 
recently used to explain the observed magnetic properties of  SrTcO$_3$~\cite{mravlje_SrTcO3_prl_2011}. 

The Mo-, Ru- and Rh- based compounds are metallic. 
Indeed, given the reduced $U/D$, it is expected and observed in
practice that oxides of the $4d$ series with a non half-filled \ttg
shell are metallic, as long as the orbital degeneracy is not too strongly lifted. 
\CROtwo, a rare example of a $4d$ $t_{2g}^4$
insulator, has indeed strong structural distortions leading to a complete orbital
polarization~\cite{gorelov_ca2ruo4_prl_2010}.
Sr$_2$MoO$_4$ and Sr$_2$RuO$_4$ are symmetrically placed with respect
to a half-filled \ttg shell, with one less and one more electron respectively, 
but their properties differ. Sr$_2$RuO$_4$ is
considerably more correlated. An orbital average of the measured
effective mass enhancements yields $m^*/m\sim 2$ for
Sr$_2$MoO$_4$ ($4t_{2g}^2$)~\cite{ikeda_Sr2MoO4_jpsj_2000} and $\sim
4$ for Sr$_2$RuO$_4$ ($4t_{2g}^4$)~\cite{imada_mit_review,mackenzie_rmp_2003}.
This distinction 
occurs because the $t_{2g}$ density of states is not particle-hole symmetric: \SROtwo has
the Fermi level close to a van Hove singularity and therefore a
smaller effective bandwidth~\cite{mravlje11}.  It is also clearly 
seen in the model calculations of Fig.~\ref{fig:Zcontour}. 

In Sr$_2$RhO$_4$ ($4t_{2g}^5$) the mass enhancement is close to
2~\cite{perry_njp_2006,baumberger06}. 
While this can be accounted for within the simple model description of Fig.~\ref{fig:Zcontour},  
recent work~\cite{martins_iridates_rhodates_prl_2011} suggests that the screened interaction 
in this compound is smaller than in the other 4d oxides, but that
the substantial renormalization comes from lifting of the degeneracy as a 
combined result of distortions and spin-orbit coupling.  

Obviously, the simple classification displayed on Fig.~\ref{fig:Zcontour} strictly applies only to 
materials in which the \ttg states are degenerate. 
It should be complemented in general with a third axis, indicating the
strength of crystal fields and other terms that lift the $t_{2g}$
orbital degeneracy. These terms, which appear due to the rotations of
the octahedra (e.g. of the GdFeO$_3$ type) and Jahn-Teller
distortions, are not negligible for all the materials considered on
Fig.~\ref{fig:Zcontour}, but the success of the classification
suggests their effects are small. In many other cases the lifting of degeneracy is
crucial, as discussed in Sec.~\ref{sec:crystal_field}. 
Spin-orbit coupling in 4d oxides reaches
0.2eV and its effects on the correlations for most of them remain to
be investigated in details in the future.

Putting all such refinements aside, the big picture is that the
materials at Janus-filling can display Hund's coupling-induced correlations while not 
being close to a Mott insulating state, and that a rich diversity of behaviour is observed 
depending on key control parameters such as filling, coupling strength, crystal-field, location of 
van Hove singularity, etc...
It is interesting in this respect to contrast the properties of ruthenates (discussed in
Sec.~\ref{sec:ruthenates}) to their rhodate equivalents which are
structurally and chemically close but have a single hole in the $4d$-shell. Unlike their Ru relatives, Rh
compounds are paramagnets: Sr$_2$RhO$_4$~\cite{perry_njp_2006} is not an unconventional
superconductor, SrRhO$_3$~\cite{yamaura_prb_2001} is not a ferromagnet and
Sr$_3$Rh$_2$O$_7$~\cite{yamaura_prb_2002} not a metamagnet with nematic behavior.

\subsection{The non Fermi-liquid `spin-freezing' regime}
\label{sec:spin_freezing}

Here, we discuss the physics of the strongly correlated metallic phase induced by the   
Hund's rule coupling, corresponding to the pale-coloured region of Fig.~\ref{fig:Zcontour}. 

Key features of this phase were pointed out in the pioneering work of Refs~\cite{werner08,haule09}.
Deep within this phase, the local moments freeze (hence the name
`spin-freezing' coined in \cite{werner08}): the local spin
susceptibility at low-temperatures increases strongly~\cite{haule09}
and the local spin-spin correlation function $\langle
S^z_i(0)S^z_i(\tau)\rangle$ does not decay at long times~\cite{werner08}.
Furthermore,  the authors of Ref.~\cite{werner08} discovered that the electronic 
self-energy at low-frequency obtained from DMFT calculations is in strong contrast to that 
of a Fermi-liquid and obeys a power-law behaviour 
$\Sigma^{\prime\prime}(\omega) \sim \Gamma + (\omega/D)^\alpha+\cdots$. 
Near the boundary of the spin-freezing regime, $\Gamma$ is small at low-T and $\alpha\simeq 1/2$. 
%
\begin{figure}[!ht]
\begin{center}
\includegraphics[width=12cm]{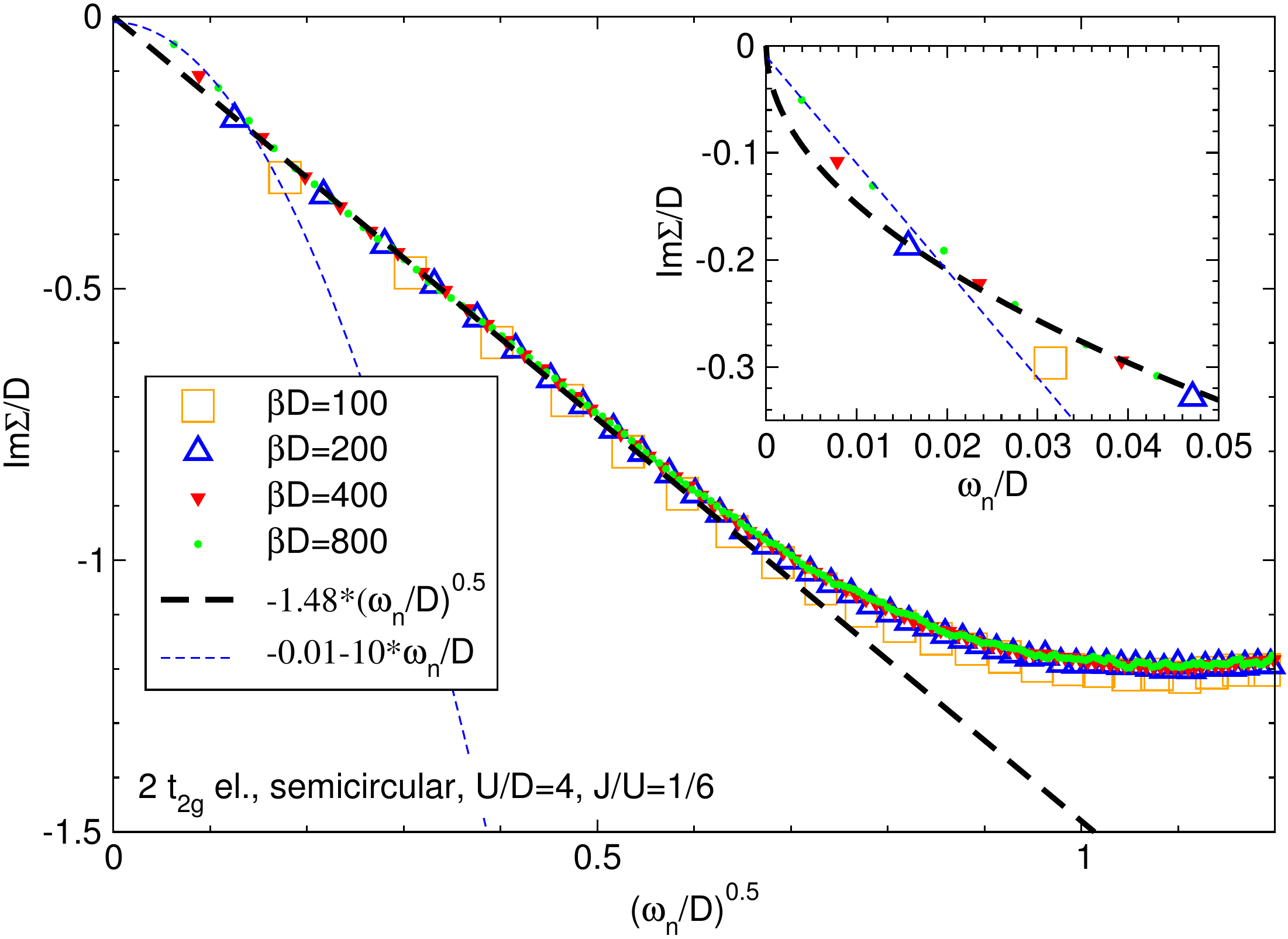}
\caption{Self-energy in the `spin-freezing' regime of the 3-orbital Hubbard-Kanamori model for 2 electrons in the band, as calculated 
by DMFT for $U/D=4$ ,  $J/U=1/6$. 
$\beta D\equiv D/kT$ is the inverse temperature normalized to the half-bandwidth. 
The plot displays $\mathrm{Im}\Sigma(i\omega_n)$ on the Matsubara frequency axis and 
emphasizes the (non Fermi-liquid) power-law behaviour $\sim (\omega/D)^{1/2}$~\cite{werner08} as well as the very low-energy crossover into a 
Fermi liquid (inset).  
\label{fig:freezing}}
\end{center}
\end{figure}
This is illustrated on Fig.~\ref{fig:freezing}, where we display the results of DMFT calculations 
on the boundary of the spin-freezing regime at `Janus' filling-factor $n=2$ and down to very 
low temperatures $T/D=1/100,\cdots,1/800$. 
These data also reveal (inset) that the power-law behaviour actually does not persist  
down to $T=0$, and that a crossover to Fermi-liquid behaviour is found for $T<\TFL$. 
The Fermi liquid scale $\TFL$ is extremely low 
however, which is another distinctive feature of this regime~\cite{haule09,mravlje11},  
and corresponds to the strong suppression of the Kondo screening scale by Hund's 
coupling discussed in Sec.~\ref{sec:kondo}.  
Note that spin-flip terms are essential in restoring Fermi liquid behaviour 
at low-temperature~\cite{Ishida_Mott_d5_nFL_Fe-SC,Biermann_nfl}. 

Besides frozen local moments, the regime $T>\TFL$ has anomalous  transport and optical properties 
which differ from that of a Fermi liquid. 
Calculations in Ref.~\cite{haule09} report a large resistivity exceeding the Mott-Ioffe-Regel criterion 
with weak temperature-dependence for temperatures much larger than $\TFL$, 
and a sharp drop upon entering the coherent regime. In the low-$T$ Fermi-liquid, a small value of the quasiparticle weight 
and a large effective mass (Fig.~\ref{fig:Zcontour}) are found. 
Non-Drude low-frequency optical response $\sigma(\omega)\sim \omega^{-0.5}$ has been 
emphasized in Ref.~\cite{werner08}. 
Many other properties of the spin-frozen regime remain 
to be worked out in details, such as a possible enhancement of the thermoelectric power. 
In Sec.~\ref{sec:ruthenates} and Sec.~\ref{sec:pnictides}, we review the implications 
of the unconventional properties of the bad-metal spin-frozen phase for the physics of ruthenates~\cite{werner08} 
and iron-based superconductors~\cite{haule09}, in connection with experimental observations. 

Finally, let us emphasize that a precise theoretical understanding of the non Fermi-liquid behaviour 
$\Sigma^{\prime\prime}(\omega) \sim \Gamma + (\omega/D)^\alpha+\cdots$ and of the 
other unconventional properties of the spin-freezing metallic regime is to a large extent an 
open and fascinating problem. 
Here, we provide a few possible hints for future work. 
The theoretical study~\cite{leo04thesis} of the relevant 3-orbital impurity problem with a local atomic hamiltonian 
(\ref{eq:ham_t2g_NSL}) has established that the low-energy $T=0$ fixed point is a Fermi liquid. 
This is clear at half-filling $n_d=3$ where the orbital fluctuations are quenched ($L=0$) and 
the low-energy effective hamiltonian is a $S=3/2$, $K=3$-channel Kondo model, which is a 
Fermi-liquid since $K=2S$. 
In contrast, for $n_d=2$ (or $n_d=4$) the angular momentum is not completely quenched in the 
$S=L=1$ 9-fold degenerate ground-state (table~\ref{table:spectrum_t2g}). 
The low-energy fixed point does remain a Fermi-liquid however 
due to the appearance of a potential scattering term in the effective hamiltonian obtained 
after eliminating high-energy states with a different valence. 
This is consistent with the observation of Fermi liquid behaviour at a very low temperature scale 
(above, and Fig.~\ref{fig:freezing}). 
It is tempting to speculate that the anomalous power-law behaviour at intermediate temperature is 
associated with a crossover controlled by a non-Fermi liquid fixed point obtained when this 
potential scattering term is absent. 
A strong-coupling analysis of the effective hamiltonian (\`a la Nozi\`eres-Blandin) indeed 
reveals~\cite{leo04thesis} a residual pseudospin-$1/2$ degree of freedom, suggesting that this fixed 
point could be related to an overscreened 3-channel, spin-$1/2$ Kondo problem. 
Another possibility is the role played by the continuous line of non Fermi-liquid fixed points~\cite{leo04thesis} 
separating the behaviour of this model (for $2\leq n_d\leq 3$) for ferromagnetic and 
`inverted' antiferromagnetic Hund's coupling. Very recently, the potential role of a ferromagnetic Kondo 
coupling emerging at low energy has also been emphasized~\cite{yin_powerlaws_arxiv_2012}. 

For another recent illustration of the potential relevance of non Fermi-liquid impurity fixed points 
to the solid-state, in the context of iron pnictides, see Ref.~\cite{ong_coleman_tetrahedron_prl_2012}.

\subsection{Spin-freezing and magnetic ordering}
\label{sec:freezing_ordering}

Another important issue is the possible development of inter-site spin correlations and magnetic long-range 
order in the spin-freezing regime. 
With such a low coherence scale for quasiparticle formation, a Doniach-type criterion would indeed suggest that this 
phase is prone to various kinds of ordering.
A very small amount of disorder could for example freeze the local moments into a phase with spin-glass order. 
Recent observations~\cite{carlo_nmat_2012} on Ca-substituted \SROtwo indeed provide experimental support to this possibility.  

Another obvious possibility for local moments in the presence of a strong Hund's coupling is ferromagnetic ordering, 
as observed e.g. in SrRuO$_3$. 
Ordering at a critical temperature higher than the low coherence scale of the paramagnetic state is  
an efficient way to restore good metallic transport. 
The direct transition from an incoherent bad metal into an ordered phase is a hallmark of 
strong correlations. It is also a major challenge to theory since, in those circumstances, ordering 
cannot be described as an instability of interacting Landau quasiparticles.
Ref.~\cite{ChanWernerMillis_3orbHub_orders} studied the magnetic and orbitally ordered phases of the 3-orbital model considered 
in this section. Although ferromagnetism is found for $U$ large enough, an extended paramagnetic spin-freezing region is 
nonetheless preserved at intermediate $U$ and filling $2\lesssim n \lesssim 4$. 

These issues deserve further studies, e.g. in the framework of cluster extensions of DMFT. 

\subsection{Competition between Hund's coupling and crystal-field splitting}
\label{sec:crystal_field}

Up to now, we have considered situations with perfect orbital degeneracy. For this reason, the general perspective 
on early TMOs provided in Sec.~\ref{sec:oxides_globalview} applies mostly to materials with only small deviations 
from perfect cubic symmetry and \ttg orbital degeneracy. 
For many materials however, it is crucial to take into account the lifting of orbital degeneracy induced by structural distortions. 
The interplay between crystal-field effects and interactions leads to a rich diversity of possible behaviours~\cite{TokuraNagaosa_OrbitalPhysics}. 
Here, we focus on the interplay with Hund's coupling. 

The key point is that the Hund's rule coupling $J$ and the crystal-field energy scale $\Delta$ 
compete with each other (see e.g. Refs~\cite{OkamotoMillis_CSRO,Lechermann_BaVS3_proceeding}). 
The former favors `orbital compensation', i.e. tends to equalize the different orbital populations so that 
the electrons distributed in all available orbitals can take full advantage of the reduction of the Coulomb 
repulsion by the intra-atomic exchange. 
The latter, in contrast, tends to populate most the lowest-lying orbitals, hence leading to `orbital polarization'. 

At a qualitative or model level, this competition can be discussed in general terms, whether the 
crystal-field splitting refers to the splitting ($10 Dq$) between \ttg and \eg states, or to the splitting between states 
within the \ttg (or \eg) manifold itself, due e.g. to a rotation of the oxygen octahedra or to a Jahn-Teller distortion of these 
octahedra.  
In practice, one should keep in mind that the order of magnitude of these two types of crystal-field 
splitting is quite different in TMOs: $1-2$~eV's for the \ttg-\eg splitting, $\lesssim 300$~meV for the intra-\ttg splitting. 

The lifting of degeneracy due to crystal-field splitting directly
affects the Mott critical coupling and hence has important
consequences for deciding whether a material is insulating or
metallic. 
The importance of this effect is best
illustrated~\cite{Pavarini_dist_d1_oxides,pavarini_njp_2005} by the
series \SVO, \CVO, \LTO, \YTO, materials which all have a nominal
$d^1$ occupancy of the \ttg shell and comparable values of $U$ and
$J$. Nevertheless, \SVO and \CVO are metals (the latter more
correlated than the former), while \LTO and \YTO are Mott insulators
(the latter with a larger gap $\sim 1\eV$ than the former $\sim
0.2\eV$).  The reason for this is the increasing orthorombic
distortion as one moves along the series (starting with cubic \SVO)
due to the 
rotations of the oxygen octahedra. 

A first effect of the crystal-field is to counteract the
effect of J on the Mott gap~\cite{demedici_MottHund}. 
In the $d^1$ case, Hund's coupling enhances $U_c$, as detailed in Sec.~\ref{sec:mottgap},
causing cubic $d^1$ oxides such as \SVO and SrNbO$_3$ to be metallic with moderate correlations. 
The crystal-field compensates this effect thus enlarging again the Mott gap and
contributing to the stronger correlations found in \CVO.

The distortion has then two further effects: it reduces the \ttg
bandwidth, and also lifts the \ttg degeneracy (by as much as $\sim 300
\meV$ in \YTO). The latter effect reduces $U_c$ (Sec.~\ref{sec:mottgap}).  Both effects increase
correlations and are responsible~\cite{Pavarini_dist_d1_oxides} for \LTO and \YTO being insulators. 
These insulators have a substantial degree of orbital polarization: for those
materials, the intra-\ttg splitting wins over Hund's coupling. 
%
For a discussion of these effects in the model context, see e.g.~Refs.\cite{manini_orbital_2002_prb,poteryaev_crystalfield_prb_2008}. 

%
BaVS$_3$, a $d^1$ material which is metallic at high-temperature is a case in which, 
in contrast, the Hund's rule coupling wins over the small ($\simeq 0.1$~eV) intra-\ttg splitting, 
leading to a compensation of orbital populations. 
This has been proposed~\cite{Lechermann_BaVS3_prl} to play a major role in explaining the development of a charge-density wave insulating state 
at low-temperature in this material. 

\begin{figure}[!ht]
\begin{center}
\includegraphics[width=12cm]{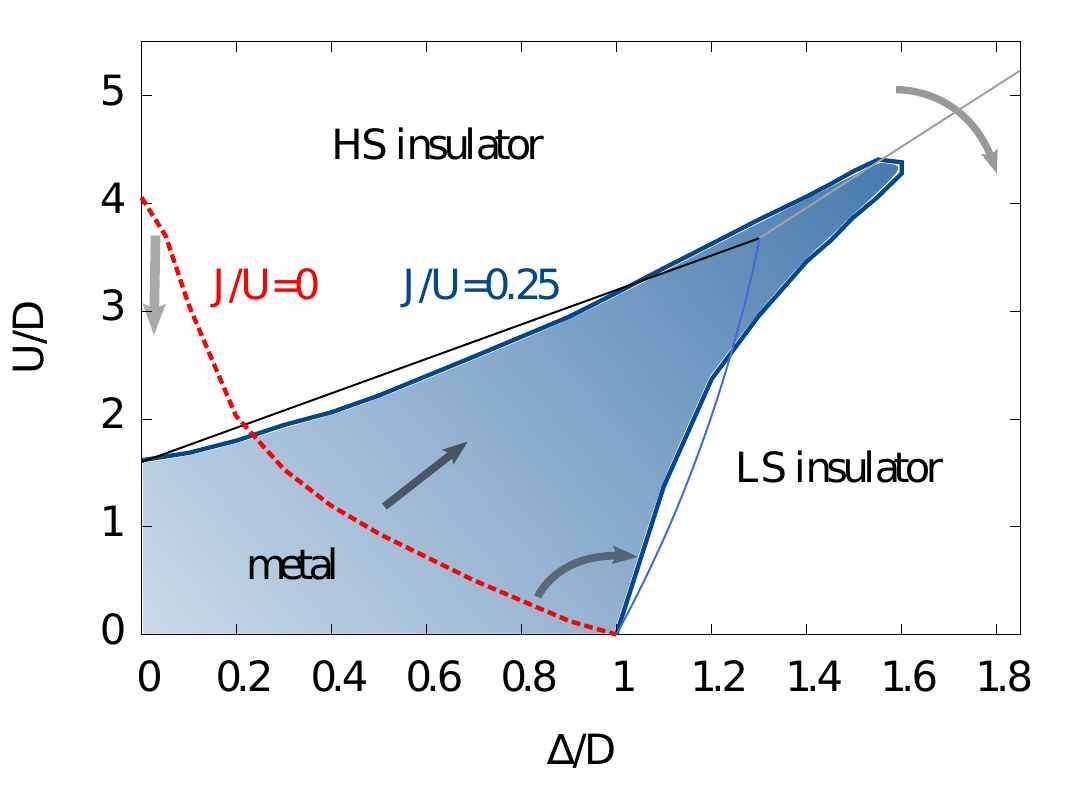}
\caption{Competition between Hund's coupling and crystal-field splitting:  
phase diagram (paramagnetic phases only) of  the two-orbital Hubbard-Kanamori model at 
half-filling, as a function of crystal-field $\Delta/D$ and interaction strength $U/D$, for a fixed value 
of $J/U=0.25$. Two insulating phases are found, one with high-spin $S=1$ (HS) and one with low-spin 
$S=0$ (LS), together with a metallic phase (in blue). 
Adapted from Ref.~\cite{werner07_hilowspin}. 
The three continuous lines denote simple estimates (see text and \cite{kunes_review_epjst_2009,kunes_HS_LS_prl_2011}) of the 
transitions between these three states, based on the atomic limit. 
Also depicted in red (dashed line) is the phase boundary separating the metallic phase (to the left) from the 
insulating phase (to the right) for $J=0$ (in this case, a LS insulator is always found except at $\Delta=0$). 
The arrows indicate how the phase boundaries move as $J$ is increased.  
\label{fig:high_low_spin}}
\end{center}
\end{figure}

The competition between Hund's coupling and the crystal-field is particularly dramatic in the strong coupling large-$U$ regime, 
where it can induce a transition between two different insulating ground-states, from high-spin (HS) 
when Hund's rule dominate to low-spin (LS) when crystal-field dominates~\cite{Bari_high-low_spin_Fe}. 
This has been the subject of several recent studies~\cite{werner07_hilowspin,kunes_MnO_natmat_2008,kunes_Fe2O3_prl_2009,
kunes_review_epjst_2009,kunes_HS_LS_prl_2011}. 
It can be simply illustrated by considering a Hubbard-Kanamori model of two bands (bandwidth $2D$) 
separated by an on-site crystal-field energy $2\Delta$~\cite{werner07_hilowspin,kunes_review_epjst_2009,kunes_HS_LS_prl_2011}. 
(For model studies of the crystal field vs. Hund competition involving three orbitals, 
see e.g.  Refs.\cite{werner09,demedici_MottHund,Kita_3band_2el_CfSplitting}). 
The generic phase diagram of this model in the half-filled case (two electrons per site) is depicted on 
Fig.~\ref{fig:high_low_spin} as a function of $\Delta/D$ and $U/D$ for a fixed ratio $J/U$. 

The transition from a HS insulator to a LS insulator can be understood from simple energetics 
in the limit of isolated atoms~\cite{kunes_review_epjst_2009,kunes_HS_LS_prl_2011}. 
We neglect first, for simplicity, the spin-flip and pair-hopping  terms in 
the Kanamori hamiltonian. 
%
At small enough crystal-field, the ground-state has $S=1$ and orbital 
isospin $T^z=0$, corresponding to each orbital occupied by one electron. The energy of this 
state is $E_{\mathrm{HS}}=U-3J-\Delta+\Delta=U-3J$. At higher crystal-field, the LS  
ground-state with two electrons in the lowest orbital ($S^z=0,T^z=-1$) has $E_{\mathrm{LS}}=U-2\Delta$. 
Hence, for $2\Delta<3J$, the HS ground-state with compensated orbital populations is favoured, 
while the LS orbitally polarized ground-state takes over for $2\Delta>3J$. 
%
%
The energies of the lowest excited states with $1$ and  $3$ electrons respectively read: 
$E_1=-\Delta$ and $E_3=U+(U-2J)+(U-3J)-2\Delta+\Delta=3U-5J-\Delta$. Hence, the Mott gap in the zero-bandwidth limit  
$\Delta_{\mathrm{at}}=E_3+E_1-2E_{\mathrm{gs}}$ reads, for density-density interactions:
\begin{equation}
\Delta_{\mathrm{at}}\,=\,U+J-2\Delta\,\,\,(\mathrm{HS}\,,\,2\Delta<3J)\,\,\,,\,\,\,
\Delta_{\mathrm{at}}\,=\,U-5J+2\Delta\,\,\,(\mathrm{LS}\,,\,2\Delta>3J)\,\,\,,\,\,\,
\label{eq:mottgap_CF}
\end{equation}
In the presence of spin-flip and pair-hopping terms, only the expression of the LS energy is modified: 
$E_{\mathrm{LS}}=U-\sqrt{(2\Delta)^2+J^2}$. The HS/LS transition occurs at $\Delta>\sqrt{2}J$~\cite{werner07_hilowspin}  and 
the atomic gap in the LS case becomes: $\Delta^{\mathrm{LS}}_{\mathrm{at}}\,=\,U-5J-2\Delta+2\sqrt{(2\Delta)^2+J^2}$. 

Hence, for a half-filled shell, one sees that when Hund's rule dominate (HS regime) the 
effective $U$ is increased  (critical $U_c$ decreased) by $J$ (and decreased by $\Delta$), 
as explained in Sec.~\ref{sec:mottgap}, while the opposite applies in the LS regime. 
By continuity with the metal to band-insulator transition at $U=0$ which occurs at $2\Delta=2D$, the phase boundary between the 
metal and the LS insulator can be approximately located by $\Delta^{\mathrm{LS}}_{\mathrm{at}}=2D$. 
The metal/HS insulator boundary can be approximated by $\Delta^{\mathrm{HS}}_{\mathrm{at}}\simeq 2D$. Note that in the 
r.h.s. of this expression, we have used the bandwidth as a measure of kinetic energy, as appropriate for $J\neq 0$ 
because of the quenching of orbital fluctuations (Sec.~\ref{sec:mottgap}), while a larger value would 
be appropriate for $J=0$. 
%
%
With these choices, the three lines separating the HS (LS) insulator and the metallic phase cross at a 
single point and yield a reasonable approximation to the phase boundaries 
calculated with DMFT (Fig.~\ref{fig:high_low_spin}, \cite{werner07_hilowspin,kunes_review_epjst_2009,kunes_HS_LS_prl_2011}). 

In the weak-coupling small $U$ regime, a calculation in the Hartree approximation~\cite{Lechermann_BaVS3_proceeding, OkamotoMillis_CSRO}
yields an effective crystal field $\Delta_{\mathrm{eff}} = \Delta + (U-5J)\delta n$, with $\delta n$ 
the orbital polarization. Correspondingly, the orbital polarizability reads 
$\chi_O=\chi_O^0/[1-c(U-5J)\chi_O^0]$. Hence, the orbital polarizability is enhanced by 
interactions if $J<U/5$ and suppressed if $J>U/5$~\cite{OkamotoMillis_CSRO,Lechermann_BaVS3_proceeding}. 
These considerations explain the slope of the 
phase boundary in Fig.~\ref{fig:high_low_spin} near the metal to LS insulator transition at small $U$. 

HS/LS transitions have been recently considered and studied in details by Kune\v{s} and coworkers 
for three materials: MnO~\cite{kunes_MnO_natmat_2008}, 
\HEM (formal valence $d^5$)~\cite{kunes_Fe2O3_prl_2009} and \LCO (formal valence $d^6$)~\cite{kunes_HS_LS_prl_2011}
along with corresponding LDA+DMFT calculations (see~\cite{kunes_review_epjst_2009} for a review). 
In both \HEM and MnO a transition is observed between a low-pressure HS insulating phase ($t_{2g}^3e_g^2$) 
and a high-pressure LS metallic phase (at $\sim 50$~GPa for \HEM and $\sim 100$~GPa for MnO). 
It is suggested~\cite{kunes_MnO_natmat_2008,kunes_Fe2O3_prl_2009,kunes_review_epjst_2009}  
that the transition in \HEM is analogous to the HS/metal transition in Fig.~\ref{fig:high_low_spin}, while 
that in MnO is more in the ionic limit, analogous to the crossing between the HS and LS atomic ground-states 
in Fig.~\ref{fig:high_low_spin} (note that 
for a $d^5$ shell, the LS ground-state is not necessarily an insulator since it has one hole in the \ttg shell).
These authors also suggest~\cite{kunes_HS_LS_prl_2011} that \LCO is an example of a material poised very close to the triple point where 
phase boundaries meet in Fig.~\ref{fig:high_low_spin}. 
In this circumstance, raising the temperature can lead to 
an entropy-driven spontaneous disproportionation with translational symmetry breaking, the HS states occupying dominantly 
one sublattice and the LS state the other.  An effective Blume-Emery-Griffiths model 
(retaining only the HS$\spinup$, HS$\spindown$ and LS states) was introduced to describe this physics. 
This electronic mechanism for disproportionation should be contrasted to the elastic (lattice) mechanism proposed in 
the early work of Bari and Sivardi\`ere~\cite{Bari_high-low_spin_Fe}. 
Both effects are likely to conspire in the actual materials. 

In this discussion, we limited ourselves to the paramagnetic phases and did not consider 
phases with long-range magnetic or orbital ordering.  
This is a vast field beyond the limited goal of this article, with a rich interplay between the Hund's rule coupling, crystal-field and structural effects, and 
superexchange and double-exchange magnetic interactions~\cite{TokuraNagaosa_OrbitalPhysics}. 
For studies of these issues at the level of the two- and three-orbital Kanamori-Hubbard models, see e.g.  Refs.~\cite{Quan_2orbitals_2el_Cfs_magnetic_order, Koyama_magOSMT, Peters_2orb_1el-orbital_magnetic_orders, Kita_CDMFT_1el_2orb_magn-orb, KoLee_SlaveRotors_2orbHub_SDW, ChanWernerMillis_3orbHub_orders}.

\subsection{Hund's coupling as a band decoupler and orbital-selective physics}  
\label{sec:osmt}

%
When orbital degeneracy is lifted by effects such as crystal-fields or different electronic structure of the bands 
(e.g. different bandwidths), correlations can affect each band in a distinct manner.
Here, we emphasize that the Hund's rule coupling enhances  such an 
orbital differentiation,  and acts, in some aspects, essentially as a `band decoupler'.

An early work which stressed that $J$ induces orbital differentiation is the
NRG study of the two-impurity composite-spin Kondo model  (\ref{eq:ham_kusu}) 
with unequal coupling strengths~\cite{yotsuhashi01}. 
For this model, the Kondo screening proceeds in two
stages~\cite{jayaprakash81,jones87}. As temperature is lowered the
system first approaches the unstable underscreened fixed point, at 
which only half of the total spin is screened, and eventually reaches the fully
screened Fermi liquid stable fixed point. 
The temperature below which full screening applies can be much reduced by $J$~\cite{jayaprakash81,yotsuhashi01}. 
The Hund's rule coupling not only suppresses both respective
Kondo temperatures, but also enhances their ratio (which can be seen
also by generalizing Eq.~\ref{eq:nevicoleman_TK} to an orbitally-dependent
case) and thereby the tendency towards orbital differentiation.

The extreme form of orbital differentiation is when the carriers on a subset of orbitals get localized,
while others remain itinerant, a concept dubbed
orbital-selective Mott phase (OSMP)~\cite{Anisimov_OSMT}.
In its simplest, almost trivial, form one can say that an OSMP is realized in 
double-exchange systems like the manganites La$_{1-x}$Sr$_x$MnO$_3$ where 
the \ttg electrons form a localized core spin, while the \eg electrons are itinerant. 

Many model studies have documented the occurrence of an OSMP  and associated 
orbital-selective Mott transition (OSMT), and that the Hund's rule coupling promotes these effects. 
The simplest model is the two-band Hubbard-Kanamori model with
unequal bandwidths $D_1$ and $D_2$, which has been thoroughly investigated (see
e.g. Ref.~\cite{Inaba_dopedOSMT} for an extensive list of
references). 
%
\begin{figure}[!ht]
\begin{center}
\includegraphics[width=15cm]{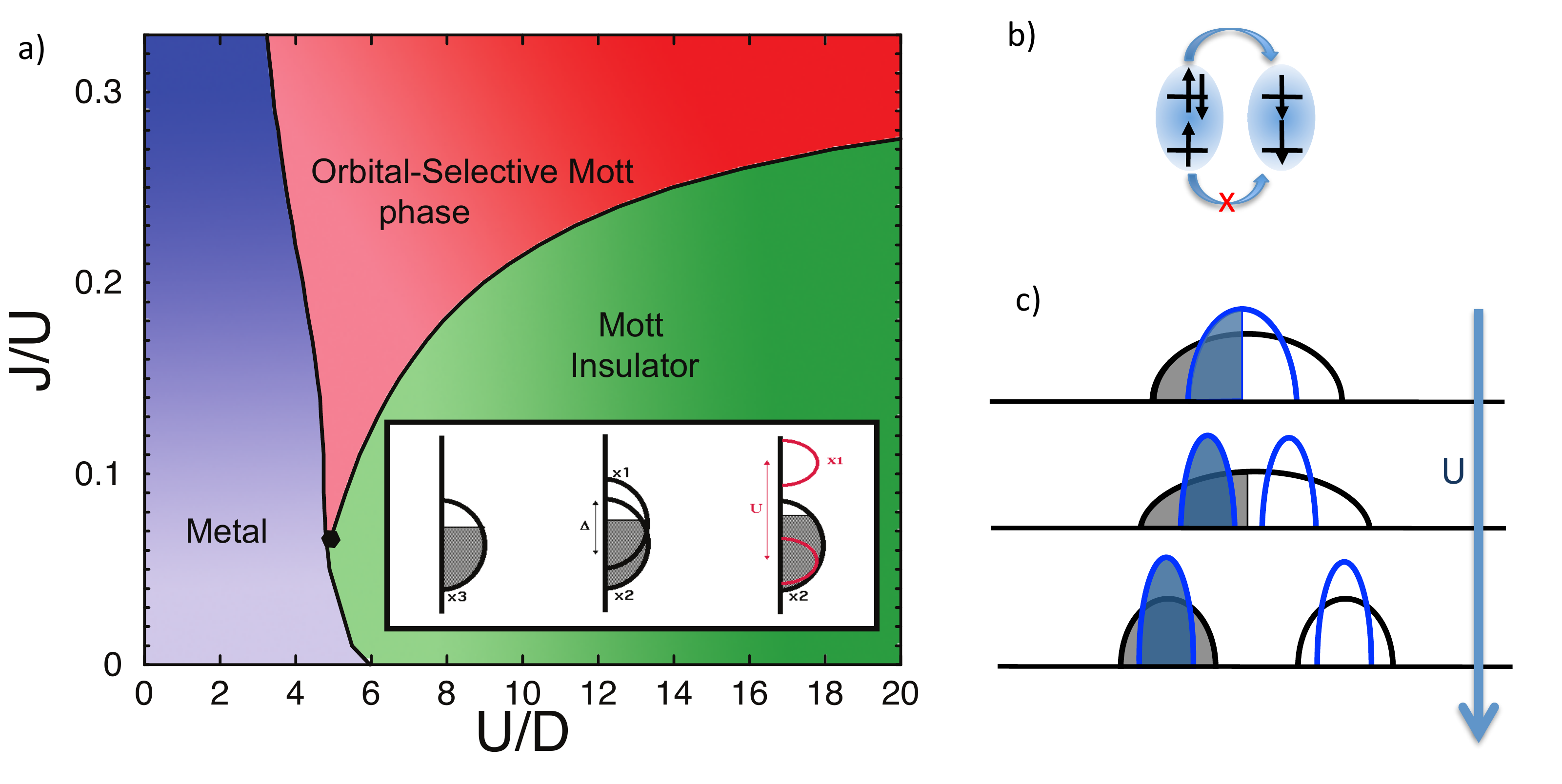}
\caption{Orbital-selective Mott physics promoted by Hund's coupling. 
(a) 
Phase diagram of a 3-band Hubbard model populated by 4
  electrons, as a function of correlation strength $U/D$ and Hund's
  coupling strength $J/U$. The crystal field lifts the 3-fold
  degeneracy so that the upper band is half-filled and the lower two
  bands that remain degenerate contain 3 electrons. An
  orbitally-selective Mott phase, in which the half-filled band has a
  gap is stabilized by $J$. Reproduced from~\cite{demedici_3bandOSMT}.
(b) Propagation of a charge excitation in two half-filled bands.  
The lower process leads to a state with energy larger by $J$ and is therefore suppressed~\cite{Koga_OSMT,Koga_OSMT_2005}.
(c) As the interaction strength in the two band model with unequal bandwidths is increased (top), 
the narrower band localizes first, and the OSMP results (middle). 
A Mott insulator (bottom) is found only at a still larger interaction strength.}
\label{fig:OSMT}
\end{center}
\end{figure}
As the correlation strength is increased (Fig.~\ref{fig:OSMT}(a) from
top to bottom), the narrower band localizes before the wider one, if
the bandwidth ratio $D_1/D_2$ is larger than a critical value.
This critical ratio is quite large $\sim 5$ for $J=0$ but already for small values of 
$J/U$ an OSMT is possible when $D_1$ and $D_2$ are of similar magnitude~\cite{demedici_Slave-spins,Ferrero_OSMT}.

An OSMT can also happen in a system of bands of the same width in
which the degeneracy is broken by the crystal field. Following
Ref.~\cite{demedici_3bandOSMT}, consider a model of 3 bands of the
same width filled with 4 electrons, with a crystal field tuned such
that there are 3 electrons in the lower two degenerate bands and the
higher band is half-filled. The half-filled band gets localized first
and as shown on Fig.~\ref{fig:OSMT}(a), a robust OSMP is found for
$J>J_c$, whose extent furthermore widens as $J$ is
increased~\cite{demedici_3bandOSMT,demedici_MottHund}.
This presumably happens because $J$ diminishes the Mott gap of the lower two
bands (Sec.~\ref{sec:mottgap}) occupied by a single hole. 
Whereas at a small $J$ the increase of $U_c$ by orbital degeneracy plays a
role~\cite{demedici_3bandOSMT}, the main effect behind this robust OSMP
is the different \emph{individual} band filling~\cite{demedici_MottHund}.
%

The relevance of individual band-filling and the importance of $J$ in promoting orbital-selective physics can be understood by recognizing that 
$J$ blocks orbital fluctuations~\cite{Koga_OSMT,Koga_OSMT_2005,demedici_3bandOSMT}. 
Koga et al.~\cite{Koga_OSMT,Koga_OSMT_2005} first noted that, for $J>0$, 
an electron added in a specific orbital cannot gain in delocalization energy
from hopping processes involving an electron in another orbital 
(Fig~\ref{fig:OSMT}b). This keeps the respective Hubbard bands and thus the Mott gaps independent. 
The OSMT then follows simply from the Mott
transitions in each individual band, which happen for distinct values of $U$ (see Fig.~\ref{fig:OSMT}c). 
The band decoupling accounts also for the behavior under doping: the OSMP
is stable~\cite{demedici_3bandOSMT,werner07_hilowspin}, until the
chemical potential exits from the widest gap (see also
\cite{Koga_OSMT,Inaba_dopedOSMT,Jakobi_dopedOSMT}).

%
The spin degrees of freedom become strongly 
inter-dependent when approaching the orbital-selective phase~\cite{greger12} however. 
Indeed, in the OSMP the system is appropriately described by a
double-exchange model and behaves as a non-Fermi liquid, due to the
scattering of the itinerant electrons on the localized
ones~\cite{Biermann_nfl, Costi_Liebsch_nFL}.

It should be noted that the model studies mentioned above aimed at unveiling 
the basic mechanism of OSMT and disregarded the possibility of a long-range ordering. 
However, at low temperature the local moments present in the OSMP carry
extensive entropy and will tend to order~\cite{Koyama_magOSMT}.
Another possibility for the system to reduce the entropy is offered by
inter-band hybridization, which can favor a singlet ground state and
replace an OSMP with a heavy Fermi-liquid at low
temperature~\cite{demedici_Slave-spins, Koga_OSMT_2005,demedici_mottpam_2005}. 
However the coherence temperature of this metallic phase will be very low if 
hybridization is small, and a selectively localized phase will be restored at finite temperature.  Likewise, even on the Fermi liquid side of the OSMT, the state at finite temperature might ressemble the OSMP.
Hence, although the occurrence of an OSMP as a stable zero-temperature phase is 
questionable, the general concept has relevance to situations in which an extended 
finite-temperature regimes with strong orbital differentiation is observed. 

To conclude our brief survey of OSMT, we turn to materials in which orbital-selective physics 
may be relevant. 
The concept of OSMT was initially proposed~\cite{Anisimov_OSMT} in order to 
explain the properties of \CSROtwo in which spin-$1/2$ local moments 
coexist with metallic transport for $x<0.5$~\cite{nakatsuji_prb_2000,nakatsuji_prl_2000,nakatsuji_prl_2003}. 
This will be discussed in more details in Sec.~\ref{sec:ruthenates_OSMT}. We shall also 
consider in Sec.\ref{sec:pnictides} the relevance of orbital-selective physics to iron-based superconductors. 

Other materials for which orbital-selective physics has been discussed are
\LiVO~\cite{anisimov_LiV2O4_prl_1999,singh_LiV2O4_prb_1999}, 
\BaVS~\cite{Lechermann_BaVS3_prl}, 
V$_2$O$_3$~\cite{Laad_V2O3,Poteryaev_V2O3},
Hg$_2$Ru$_2$O$_7$~\cite{Craco_Hg2Ru2O7} and CoO under
pressure~\cite{Huang_CoO_OSMT}. 
Following an early suggestion of Goodenough~\cite{goodenough_transition_metals_pr_1960}, 
recent LDA+DMFT studies of elemental metallic $\alpha$-Fe~\cite{Katanin_Iron} suggest that the 
d-electrons in $t_{2g}$ bands are itinerant, in contrast to the ones in $e_g$ bands
which form local moments due to Hund's exchange.  A similar situation was proposed for FeO
under pressure~\cite{Shorikov_FeO}, but a different result (a
high- to low- spin crossover) is reported from fully charge self-consistent LDA+DMFT calculations~\cite{Ohta_FeO}. 
Finally, the relevance of the OSMT concept to heavy-fermion physics was recently discussed and reviewed 
by M.~Vojta~\cite{vojta_OSMT_heavy_fermions_review_jltp_2010}. 

\section{Ruthenates}
\label{sec:ruthenates}

In this section we give a brief overview on perovskite ruthenates
A$_{n+1}$Ru$_n$O$_{3n+1}$ with A being Ca or Sr. In these materials $4$ 
electrons occupy the three t$_{2g}$ orbitals. Compared to the $3d$ TMOs, 
the extended nature of $4d$ orbitals gives rise
to moderate values of the screened interaction $U \sim 2$eV and broad
bands $W=2D\sim 3\mathrm{eV}$. The larger overlap of $4d$ orbitals with
oxygens enhances the \ttg- \eg crystal field splitting. Hence, a  
high-spin state is not realized here in contrast to the isoelectronic $3d$ LaMnO$_3$.  
In spite of $U\lesssim W$ and 3-fold t$_{2g}$ orbital
degeneracy, these materials are quite correlated with specific heat
enhancements $\gamma/\gamma_\mathrm{LDA}>4$.

\subsection{Ruthenathes in a nutshell}
\label{sec:ruthenates_overview}

The properties of a few widely-investigated ruthenates are listed in Table~\ref{table:ruthenates}. 
\begin{table}[!b]
\begin{center}
\begin{tabular}{c c c c c }
Compound & Magnetic order &  $\gamma/\gamma_{\rm{LDA}}$ &  $\rho \propto T^2$ &  Remarks \\
\hline
Sr$_2$RuO$_4$ & PM & 4 & $<25$\,K &   unconv. SC $<1.5$\,K \\
SrRuO$_3$ & FM $<160$\,K & 4 & $<15$\,K &   $\sigma \propto \omega^{-0.5}$    \\
Sr$_3$Ru$_2$O$_7$ & PM &  10 & $<10$\,K  &  MM QCP and nematicity  \\ 
CaRuO$_3$ & PM & 7 &  $T^{1.5}$ $>2$\,K &  $\sigma \propto \omega^{-0.5}$, $\gamma=\gamma_{\mathrm{FL}}+\log(T)$\\
Ca$_2$RuO$_4$ & AF $<110$\,K  & \ding{55} & \ding{55} &  insulator $<310$\,K \\
\end{tabular}
\caption{\label{table:ruthenates} Ruthenates in a nutshell. }
\end{center}
\end{table}

We start by the single-layer $n=1$ compound. Sr$_2$RuO$_4$ has a 
body-centered tetragonal unit cell.  Below $T_c=1.5$K it becomes
superconducting. The unconventional superconductivity in a material
isostructural with LSCO cuprates generated wide interest. The
superconductivity and normal state properties are reviewed
in~\cite{mackenzie_rmp_2003,bergemann_adv_phys_2003}. Above $T_c$, 
Sr$_2$RuO$_4$ is a paramagnetic metal with Fermi-liquid behavior at
low temperatures. The carrier masses are enhanced 
with $\gamma/\gamma_\mathrm{LDA} \approx 4$. Despite the (small) tetragonal
splitting, $4/3$ of an electron is found in each of the 
orbitals\footnote{In LDA a slight polarization in favor of xz-yz
orbitals is found. The discrepancy between theory and
quantum-oscillation experiment~\cite{mackenzie_prb_1998} is diminished
if the atomic physics (Hund's coupling) is treated appropriately,
such as in LDA+DMFT~\cite{mravlje11}.}.

The 3-dimensional SrRuO$_3$ is an itinerant ferromagnet with Curie
temperature $T_c=160$\,K (see \cite{koster_SrRuO3_rmp_2012} for a recent review). 
It crystallizes in a rhombohedral GdFeO$_3$
structure, in which the octahedra are tilted by $\sim 10$~degrees
from an ideal cubic structure, see, e.g.~\cite{zayak_prb_06}. 
Optical spectroscopy revealed $\mathrm{Re}\, \sigma(\omega) \propto
{\omega}^{-0.5}$~\cite{kostic_prl_1998,dodge_prl_2000}.  Despite this
anomalous dependence, at low temperatures quantum
oscillations~\cite{mackenzie_prb_1998} and strict $T^2$ resistivity
below 15K have been found \cite{capogna_prl_2002}. Specific heat
enhancements $\gamma/\gamma_\mathrm{LDA}=3.7$ \cite{allen_prb_96} and
$4.4$~\cite{okamoto_prb_99} have been reported.

The bi-layer compound Sr$_3$Ru$_2$O$_7$ is a paramagnetic metal.  It
is situated very close to the metamagnetic quantum critical point
which is reached upon applying a magnetic field of $7.9\,$~Tesla along the
c-axis \cite{grigera_science_2001}. At very low temperatures, an electronic 
nematic state forms (see \cite{fradkin_annrev_2010} for a review). 
The carrier masses are strongly enhanced, with $\gamma/\gamma_\mathrm{LDA} \sim 10$ at zero-field. 
A $T^2$ resistivity is observed below 7K~\cite{capogna_prl_2002}.

Table~\ref{table:ruthenates} contains also two Ca- substituted
ruthenathes.  The smaller Ca ion causes a stronger distortion of the
lattice. The infinite-layer compound CaRuO$_3$ has a stronger
rhombohedral distortion than SrRuO$_3$ with octahedra tilted by $17$
degrees \cite{zayak_prb_06}, is paramagnetic and has a large
$\gamma=74$\,mJ/molK$^2$ \cite{shepard_jap_1997} corresponding to an
enhancement $\sim 7$ over LDA value.  Compared to SrRuO$_3$ the mass
enhancement is larger most likely because CaRuO$_3$ is not
ferromagnetic.  Similar anomalous dependence of optical conductivity
as in the Sr- compound is found \cite{lee_prb_2001,lee_prb_2002}. Down
to a few Kelvin $\rho \propto T^{1.5}$ \cite{capogna_prl_2002}.

Ca$_2$RuO$_4$ is the only insulating ruthenate. Following a structural
distortion, it becomes insulating below 365K~\cite{alexander_prb_1999}
and orders antiferromagnetically below
110K~\cite{cao_prb_ca2ruo4_1997}.  The insulating state has been
explained~\cite{gorelov_ca2ruo4_prl_2010} in terms of the complete filling of the xy orbital which occurs
due to the compression of oxygen octahedra along the
c-axis in the low-temperature S-Pbca
structure, followed by a transition to a Mott insulator which occurs
in the narrower bands spanned by the $xz,yz$ orbitals with $W<U$.  The
phase-boundary can be shifted by application of
pressure~\cite{steffens_prb_2005}. Interestingly, upon substituting a 
few percent of Ru for Cr, a negative thermal expansion is
found~\cite{qi_prl_2010}. 

\subsection{Origin of correlations}
\label{sec:ruthenates_why_correlated}

\begin{figure}[!ht]
\begin{center}
\includegraphics[width=15cm]{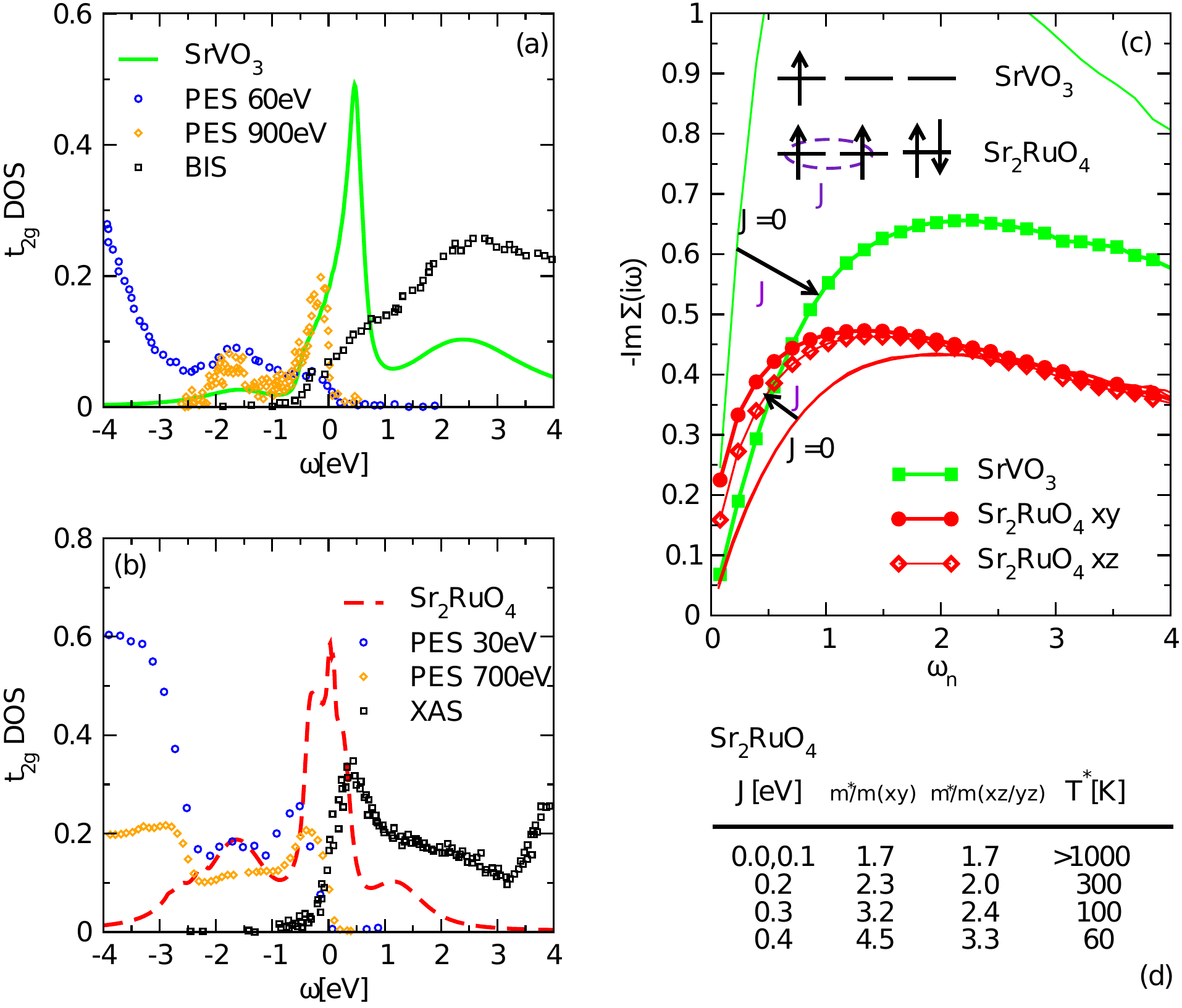} 
\caption{
Strong correlations from Hund's coupling in ruthenates. 
(a) The SrVO$_3$ $t_{2g}$ LDA+DMFT density-of-states (DOS) compared to the results of the X-ray
photoemission~(PES)~\cite{morikawa_prb_1995_pes_bis_srvo3_cavo3,sekiyama_prl_2004_high_resolution_pes_srvo3}
and inverse-photoemission  
(BIS)~\cite{morikawa_prb_1995_pes_bis_srvo3_cavo3}.
High energy PES \cite{sekiyama_prl_2004_high_resolution_pes_srvo3} is
more sensitive to the d-states and resolves better the quasi-particle DOS. 
(b) The Sr$_2$RuO$_4$ LDA+DMFT density-of-states compared to the valence-band
PES from Ref.~\cite{okuda_98_pes}, high-energy
PES~\cite{pchelkina_prb_2007} and X-ray absorption spectroscopy
(XAS) \cite{kurmaev_98_xas}. 
(c) Imaginary part of the Matsubara self-energies. The results at the physical values of the interactions
$U=2.3$eV, $J=0.4$eV for Sr$_2$RuO$_4$ and $U=4.5$eV, $J=0.6$eV for
SrVO$_3$ are compared also to the results with the same $U$ but
$J=0$. 
(d) Table (from Ref.~\cite{mravlje11}) displaying the mass enhancements $m^{*}/m_{\mathrm{LDA}}=
Z^{-1}=1-\partial \Sigma(z)/\partial z |_{ z=0^+}$ for each orbital. 
The coherence temperature $T^*$ is defined as the highest
temperature where $Z \mathrm{Im} \Sigma(0+) \le k T$ holds for both orbitals. 
\label{fig:pes_sr2ruo4_srvo3}}
\end{center}
\end{figure}

Overall, the ruthenates exhibit several remarkable properties signalling a correlated metallic state, 
with the carrier masses significantly enhanced over the LDA predictions. 
Where do the strong correlations come from ? 

In several 3d oxides, the strong correlations appear due to the proximity to a
Mott insulating state, as revealed  e.g. by the pronounced Hubbard bands
observed in photoemission spectroscopy.  
On Fig.~\ref{fig:pes_sr2ruo4_srvo3} we plot the LDA+DMFT $t_{2g}$
density-of-states for a 3d oxide SrVO$_3$, and compare it to the data
from (inverse-) photoemission spectroscopy. The data show the
quasi-particle band (visible to a lesser extent in the inverse
photoemission and low-energy photoemission) and also the signatures of
the Hubbard bands. Whereas the upper Hubbard band overlaps also with the
$e_g$ states and can thus not be identified unambiguously, the oxygen
contribution to the spectra is easily identifiable (large peak below 3eV)
and has been subtracted out from the data in Ref.~\cite{sekiyama_prl_2004_high_resolution_pes_srvo3}. 

On Fig.~\ref{fig:pes_sr2ruo4_srvo3}(b) the data is plotted for
Sr$_2$RuO$_4$. Encouraging agreement with experiment is also found there. 
Comparing the two materials, one sees that the Hubbard bands
have a larger separation in the case of SrVO$_3$, corresponding to the
larger value of the interaction for this compound. 
The peak-to-peak distance between the Hubbard bands in the two
compound differs by an amount corresponding to the respective
$U_\mathrm{eff}=U-3J$ values.

Fig.~\ref{fig:pes_sr2ruo4_srvo3}(c) displays the imaginary part of the LDA+DMFT 
self-energies on the Matsubara axis $\omega_n=(2n+1)\pi kT$.  
For SrVO$_3$, the larger $U/W$ induces large values
of $\mathrm{Im}\Sigma(i\omega_n)$ at large frequencies. At smaller frequencies well
defined quasi-particles are rapidly recovered: the data points  
are linearly aligned and intercept the y-axis with a small slope
(corresponding to $Z\sim 0.5$) and at a small value corresponding to a scattering rate  
$\Gamma=\mathrm{Im}\Sigma(i0^+) \ll kT$. 
In contrast, Sr$_2$RuO$_4$ displays weaker correlations (smaller $|Im \Sigma|$ 
) 
at high frequency, but those correlations turn stronger at  low frequency, 
giving rise to a large slope corresponding to $Z=0.2$ for the $xy$ and $Z=0.3$ for the
$xz$ orbital. 
Note that the correlations are weaker for the $xz,yz$ orbitals in spite of their {\it smaller} bandwidth 
(which is therefore not a crucial physical ingredient here). 
Indeed, quantum oscillations experiments reveal that the largest mass renormalization 
$\sim 5$ correspond, surprisingly \cite{konik_prb_2007}, to the widest $xy$ band. 

The persistence of correlations to low energies in Ru- but not in the 
V-compound is suggestive of the Hund's rule coupling. This binds a
pair of holes on a Ru-ion into a high spin (Table~\ref{table:spectrum_t2g}), 
but does not affect the single-electron ground state multiplet of the $n=1$ SrVO$_3$ compound. 
On Fig~\ref{fig:pes_sr2ruo4_srvo3}(c) we also show the LDA+DMFT 
results for $J=0$. 
For SrVO$_3$, suppressing $J$ {\it increases} correlations at all frequencies, and brings 
the material in proximity to a Mott insulating state. Indeed, at $U_\mathrm{eff}=5$eV,
a Mott insulator is found within a \ttg description.  
In contrast, for Sr$_2$RuO$_4$ setting $J=0$ does not influence much 
the correlations at higher energies in spite of the increased $U_\mathrm{eff}$. 
On the other hand, the low frequency correlations disappear.
Such behavior is found also in LDA+DMFT calculations for other ruthenathes, thus indicating that 
the strong correlations in these compounds are due to the Hund's rule coupling.

\subsection{Physical consequences of correlations induced by the Hund's rule coupling}
\label{sec:ruthenates_physics}

\subsubsection{Coherence-incoherence crossover in Sr$_2$RuO$_4$}

Together with the large mass enhancements, the scale below which Fermi liquid behaviour 
applies is found to be quite low in \SROtwo. 
The crossover out of the Fermi liquid is seen by
several experimental probes.  Despite a large anisotropy (with
$\rho_c/\rho_{ab}>1000$ at low $T$), the in-plane $\rho_{ab}$ and
out-of-plane $\rho_c$ resistivity both initially increase as $T^2$ up
to $T_\mathrm{FL}=25K$ \cite{hussey_prb_1998}. 
At a temperature 130K $\rho_c$ reaches a maximum and diminishes if the
temperature is raised further. Conversely $\rho_{ab}$ retains 
metallic dependence and increases up to the highest temperature (1300K)
measured \cite{tyler_prb_1998} without any sign of saturation.  In
addition to transport, ARPES \cite{wang_prl_2004, kidd_prl_2005} and
NMR \cite{imai_prl_1998} also reveal a low coherence scale. In ARPES
quasi-particles persist up to 150\,K, in NMR Korringa law
$1/T_1 \propto T$ is seen only below 50\,K. 

A theoretical calculation within LDA+DMFT~\cite{mravlje11} has
accounted for many aspects of the experiments. 
A coherence scale $T^*$ was defined by comparing the inverse quasiparticle lifetime to 
$kT$, and the Hund's coupling was shown to be essential in explaining the low value of $T^*$ 
(Fig.~\ref{fig:pes_sr2ruo4_srvo3}). 
%
Quantitative agreement with ARPES and NMR was found. Curves shown on
Fig.~\ref{fig:pes_sr2ruo4_srvo3} are based on the unpublished data
from that work.  The table, Fig.~\ref{fig:pes_sr2ruo4_srvo3}(d)
displays the mass renormalizations and coherence scale as a function of
$J$.  A larger mass renormalization is found for the $xy$ orbital ($\gamma$-band), 
in agreement with experiment. This has been related to
the proximity to the van-Hove singularity in the $xy$ band. 
Note that this differentiation between the $xy$ and $xz,yz$ bands occurs only once the
Hund's rule coupling is turned on, due to the 'orbital-decoupling'
action of $J$, discussed elsewhere in this review (Sec.~\ref{sec:osmt}). The proximity to a
van-Hove singularity thus cooperates with $J$ to make ruthenates strongly
correlated materials, despite their small $U/W$-ratio.

\subsubsection{Non-Fermi-liquid behavior in SrRuO$_3$ and CaRuO$_3$.}
\label{sec:ruthenates_NFL}

\begin{figure}[!ht]
\begin{center}
\includegraphics[width=15cm]{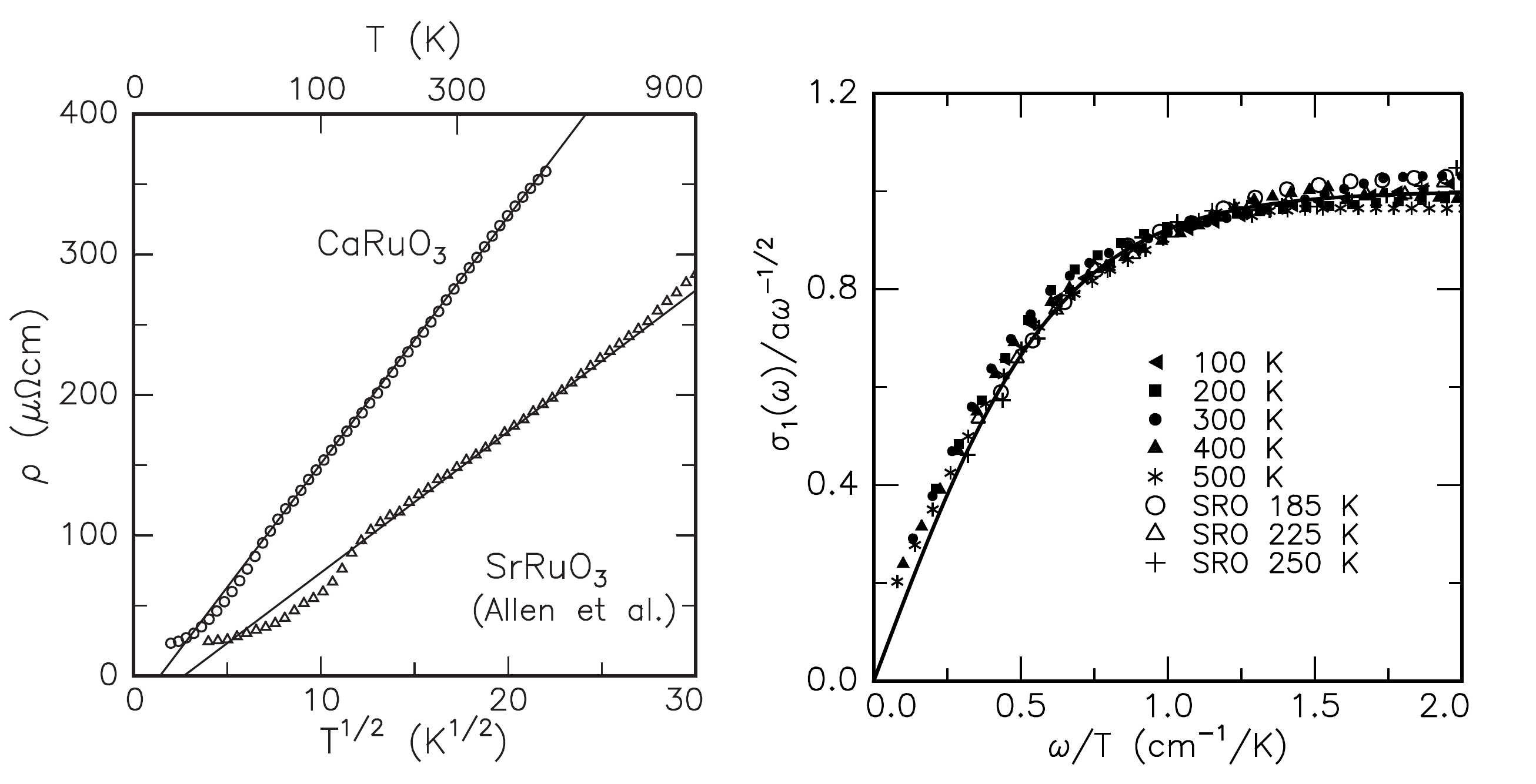}
\caption{Non Fermi Liquid behaviour in SrRuO$_3$ and CaRuO$_3$, possibly related to 
Hund's coupling physics and `spin freezing'. 
Left panel: resisivity vs. $T^{0.5}$. Right panel: Optical conductivity showing $1/\omega^{1/2}$ behaviour and $\omega/T$ scaling.   
From Ref.~\cite{lee_prb_2002}. 
\label{fig:lee_scaling}}
\end{center}
\end{figure}

In ruthenates, the resistivity at very high temperatures exceeds
\cite{cao_sol_st_comm_2004} the Mott-Ioffe-Regel limit.  Nevertheless,
at low temperatures $T<\TFL$ electrons in ruthenates form a
Fermi-liquid. The signatures of the Fermi-liquid behavior such as the
observation of quantum oscillations and the $T^2$ law in resistivity
has by now been seen in all metallic ruthenates, very recently also on
thin-film samples of \CRO below $2$K, measured in P.~Gegenwart's group
at the time this article is being
written (M.~Schneider et al., unpublished). 
The Fermi-liquid temperature $\TFL$, on the other hand, is quite small, 
and the ruthenates provide a tantalizing ground
for trying to identify their behaviour for $T>\TFL$ in terms of a universal but
non-Fermi liquid regime. 
So far, the most successful such identification has been in the measurements 
of optical conductivity in SrRuO$_3$ and
CaRuO$_3$ \cite{kostic_prl_1998,dodge_prl_2000,lee_prb_2002,koster_SrRuO3_rmp_2012}.
Fig.~\ref{fig:lee_scaling} from Ref.~\cite{lee_prb_2002} summarizes this
data. It shows that $\omega/T$ scaling applies and that
the optical conductivity at large enough frequencies obeys 
$\sigma_1(\omega)\sim \omega^{-1/2}$.
Another signature of the non-Fermi liquid, which is seen in
CaRuO$_3$~\cite{cao_sol_st_comm_08} and in Ca substituted
Sr$_2$RuO$_4$~\cite{nakatsuji_prl_2003} is a $\log(T)$ correction to
$C/T$.  The origin of this has not been clarified yet. In particular
it remains to be shown whether it is an intrinsic property of the
correlated state with a low-coherence scale.  We notice that the few
lowest temperature data points of Ref.~\cite{cao_sol_st_comm_08}
display saturation of $C/T$ and may be indicative of the eventual
formation of a Fermi liquid below 3K.
 
 Overall, it is quite tempting to associate~\cite{werner08} the NFL regimes observed in \SRO and \CRO 
 to the power-laws found in the `spin-freezing' regime for $T>\TFL$ discussed 
 in Sec.~\ref{sec:spin_freezing}. Obviously, this fascinating possibility deserves further 
 investigations.

\subsubsection{Ca${_{2-x}}$ Sr$_{x}$ RuO$_4$, heavy carriers and orbital selectivity.}
\label{sec:ruthenates_OSMT}

Partial substitution of Ca into Sr$_2$RuO$_4$ leads to a rich physics and phase diagram~\cite{nakatsuji_prl_2003}. 
%
The smaller size of Ca induces rotations of octahedra which
appear first at $x=1.5$ and progressively become more pronounced with
diminishing $x$ until reaching almost 13 degrees at $x=0.5$. 
A strong ferromagnetic enhancement of the magnetic susceptibility with Curie-Weiss
behaviour corresponding to an $S=1/2$ moment is found in a wide range $0.2\lesssim x \lesssim 1.5$~\cite{nakatsuji_prb_2000}. 
Note that one would expect an $S=1$ moment for an isolated Ru atom with 4 electrons. 
At $0.2<x<0.5$ a weak rhombohedral distortion appears ~\cite{steffens_prb_2011}.
%
%
Below $x=0.2$, stronger rhombohedral distortions with compressed octahedra lead to 
an insulating state, see discussion on Ca$_2$RuO$_4$ above. 

Especially interesting is the regime close to the structural transition at
$x=0.5$. The coexistence of metallic transport with an $S=1/2$. 
Curie-Weiss magnetic susceptibility has inspired
Anisimov et al. to propose that an orbitally-selective Mott transition (OSMT) occurs~\cite{Anisimov_OSMT}. 
In this scenario,  $1/3$ of an electron would be transferred from the metallic $xy$ band, 
and the 3 electrons in the narrower $xz,yz$ bands would localize. 
However, there is by now much experimental evidence against this proposal, 
the most direct being the ARPES observation of all three Fermi
surface sheets~\cite{wang_prl_csro_2004}. 
The unchanged-position of the nesting-induced peaks at incommensurate wave-vectors 
in the susceptibility~\cite{steffens_prb_2011} also suggest that the charge-transfer does not occur. 

In fact, it is the $xy-$ band that displays the strongest correlations and the 
heaviest carriers. This is already the case for Sr$_2$RuO$_4$, as discussed above. 
With diminishing $x$, the correlations gradually become
stronger, as evidenced by the decrease of $T_\mathrm{FL}$ 
(identified as the scale below which $\rho \propto T^2$) and by the increase of the
specific heat coefficient $\gamma$. Close to $x=0.5$ the carriers
become very heavy, with $\gamma\sim 250$\,$\mathrm{mJ/molK^2}$, about
20 times the LDA value. The optical spectroscopy
data~\cite{lee_prl_2006} points at a mass enhancement associated mainly with 
the $xy$ band. Similar indications follow from the polarized neutron diffraction study at $x=0.5$, 
which found that, in the presence of a magnetic field, the moment is on the 
$xy$-orbitals and the adjacent oxygen
sites~\cite{gukasov_prl_2002}. ARPES data at $x=0.2$ is controversial,
one study reporting all the Fermi sheets~\cite{shimoyamada_prl_2009}
whereas another study does not see the $xy$ sheet~\cite{neupane_prl_2009}.

In our view, a possible comprehensive explanation of this rich behaviour is reached
by recognizing that the effects of the Hund's rule coupling and of the
proximity to a van-Hove singularity, responsible for heavy carriers
and orbital differentiation in the undistorted Sr$_2$RuO$_4$, become
amplified by structural distortions in Ca$_{2-x}$ Sr$_{x}$RuO$_4$. 
Certainly, the value of $J$ does not change upon rotations of the octahedra, 
however the effective band-widths do.  Indeed, the dominant effect is the 
narrowing of the band originating from the in-plane $xy$-orbitals~\cite{fang_prb_2001}. 
The effects of $J$ on the electrons with lower Fermi velocity, and its 'band-decoupling' action leads to poorly
screened moments on $xy$ orbitals and incoherent carriers. This accounts for the $S=1/2$ Curie-Weiss
susceptibility even though strict OSMT may not occur. In fact, at
higher temperatures, a $S=1/2$ moment is observed in an extended range
$0.2<x<1.5$. Below $x=0.5$ when the octahedra also tilt, the
$xz,yz$-derived bands narrow down and the corresponding correlations
increase, as perhaps indicated by the build-up of incommensurate
magnetic fluctuations.
These qualitative ideas call for a detailed study using LDA+DMFT techniques. 

The poorly screened moments are susceptible to ordering at low temperature. 
Close to the $x=0.5$ critical point, Nakatsuji et
al~\cite{nakatsuji_prl_2003} found a history-dependent magnetization
compatible with the build-up of short-range ferromagnetic ordering. The
phase diagram has very recently been refined in a $\mu SR$ study,
which revealed subtle signatures of spin-glass ordering with moments
below $0.2$\,$\mu_B$ at most Ca concentrations~\cite{carlo_nmat_2012}.

Finally, in the Ca-rich region $0<x<0.2$, antiferromagnetic insulator is
found, with properties that of the $x=0$ end-compound Ca$_2$RuO$_4$
discussed above. The metal-insulator transition coincides with the
structural transition from L-Pbca to S-Pbca. The transition
temperature diminishes with increasing $x$ and vanishes a bit below
$x=0.2$. Only the rotations and tilts are not sufficient to turn a
ruthenate insulating, a compression of the octahedra realized in the
S-Pbca phase which completely polarizes the orbitals is needed.

\section{Iron-based superconductors as Hund's correlated metals}
\label{sec:pnictides}
%
The recent discovery of high-temperature superconductivity~\cite{Kamihara_pnictides1,Kamihara_pnictides2} in
iron pnictides and chalcogenides, has generated considerable interest 
(for reviews see e.g. Refs.~\cite{Hu_Li_AnnualRev_FeSC,stewart_ironSC_rmp_2011}).  
Obviously in the limited space of this article we will not attempt to cover the intensive research performed on the subject.
We will rather focus on the importance of the Hund's coupling for the physics of these compounds.

Right from their discovery, the degree of electronic correlation in these materials has been debated, with views ranging 
from the itinerant limit with magnetic correlations
induced by nesting~\cite{Mazin_Splusminus,Xu_Ising_Pnictides}, all the
way to localized magnetism~\cite{Si_Abrahams_J1J2}.
The importance of electronic correlations while keeping a metallic
description has been emphasized early on in
Refs.~\cite{Haule_LaOFFeAs_PRL2008} and \cite{haule09}.

In our view, it has now become clear that these materials do
display important effects of electronic correlations. From a
phenomenological standpoint (looking for example at the Drude weight,
specific heat enhancement, renormalization of bandwidth and Fermi
velocities, etc.), the degree of correlation clearly increases when
going over the different materials, in the order (from weaker to
stronger correlation effects): 1111 pnictides (such as LaFeAsO), 122
(such as \BFA), 111 (such as LiFeAs) and, at the more strongly
correlated end~\cite{tamai_baumberger_chalcogenides_prl_2010}, the 11 chalcogenides (FeSe, FeTe).
An issue which is still controversial is whether these differences are mainly due to variations in
the structural properties with similar interaction strengths
($F^0 , J_H$) \cite{Yin_kinetic_frustration_allFeSC,Kutepov_pnictides_parameters-GW} or whether it is
essential to take into account an increase of the interactions, especially for the 11 chalcogenides
\cite{miyake_interactions_jpsj_2010,miyake_pnictides_jpsj_2008,
aichhorn09,Aichhorn_FeSe,Ishida_Mott_d5_nFL_Fe-SC,Liebsch_FeSe_spinfreezing}. 
By and large, the big picture is nonetheless that the correlations are important.

\begin{figure}[!ht]
\begin{center}
\includegraphics[width=15cm]{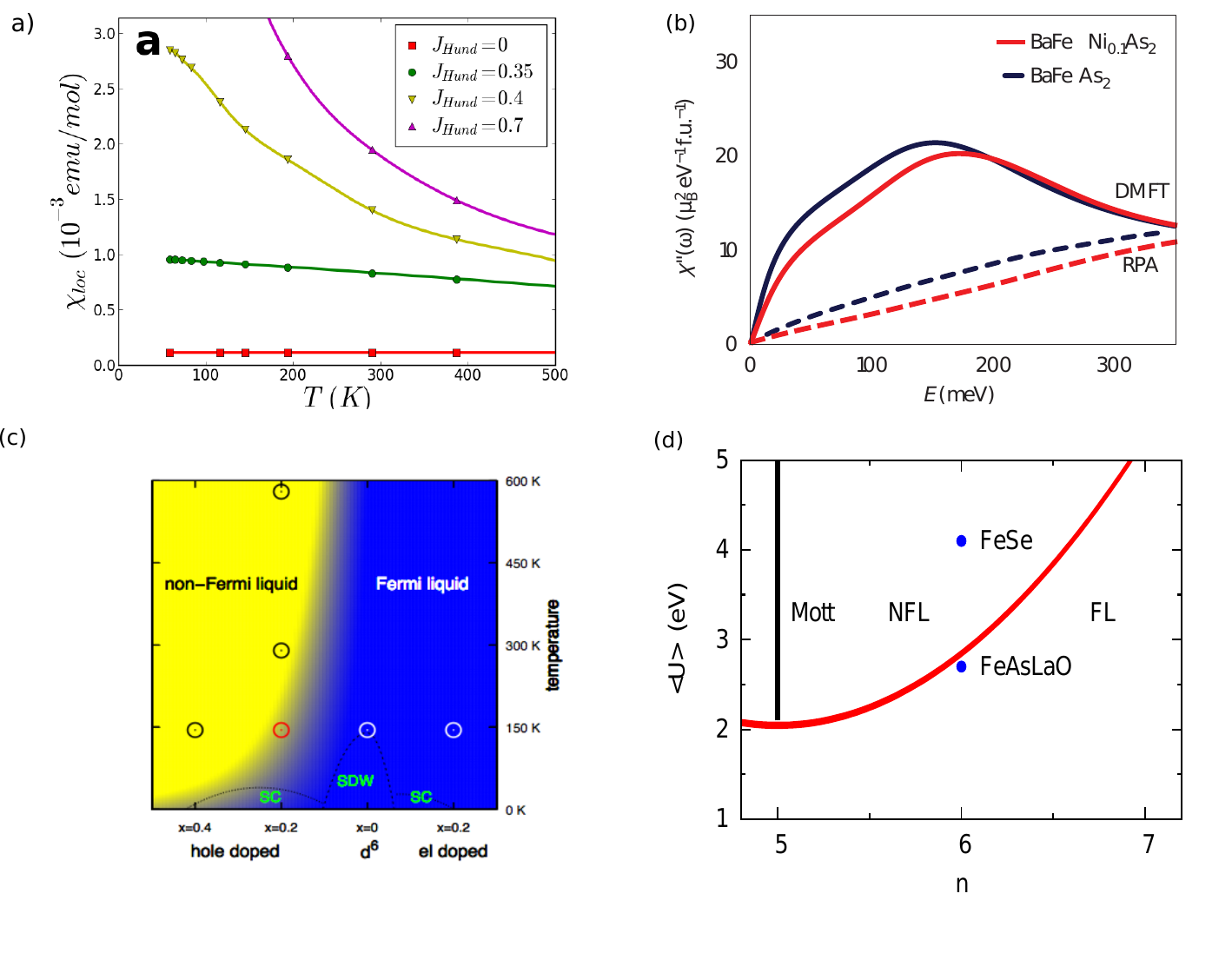}
\caption{Iron-based superconductors as `Hund's metals'. 
{\bf a.} Temperature-dependence of the local susceptibility for a 5-band LDA+DMFT description of 
LaO$_{1-x}$F$_x$FeAs, revealing the sensitivity to Hund's coupling (from Ref. \cite{haule09}). 
{\bf b.} RPA and LDA+DMFT calculations of  $\chiloc^{\prime\prime}(\omega)$ in absolute units 
for BaFe$_2$As$_2$ and BaFe$_{1.9}$Ni$_{0.1}$As$_2$ (from Ref. \cite{Liu_magmoments_122}). 
{\bf c.-d.} Spin-freezing region with power-law non Fermi-liquid (NFL) self-energy:   
({\bf c}) For doped \BFA , as obtained in the LDA+DMFT  study of Ref. \cite{{Werner_122_dynU}},  
({\bf d}) Schematic boundary in the $U$ vs. filling diagram (from Ref.\cite{Liebsch_FeSe_spinfreezing}), 
illustrating the stronger correlations in the chalcogenides.
\label{fig:pnictides}}
\end{center}
\end{figure}

The key role of the Hund's coupling has been recognized early on for these materials. 
In a pioneering article, Haule and Kotliar~\cite{haule09} proposed that Hund's
coupling may indeed be responsible for the correlation effects and thus for the unconventional aspects of the metallic state.
Within 5-bands LDA+DMFT calculations, they found that the Hund's
coupling dramatically reduces the coherence scale $T^*$ below which a
metal with Pauli susceptibility is found, leaving an incoherent metal
with local moments for $T>T^*$ (see Fig.\ref{fig:pnictides}a).
It was also recognized early on \cite{Johannes_local-moment}
that the Hund's coupling is responsible for the formation of the iron-local moment 
in these compounds.  
This is consistent with X-ray spectroscopy~\cite{Yang_Devereaux_weakcorr}
which reported 
a large value of $J\sim 0.8$\,eV .

%
%
Unexpectedly, in the magnetic state, the LSDA was found to overestimate
the size of the ordered magnetic moments ($\sim 2\mu_B$, whereas
experiments yield  moments $<1\,\mu_B$). The LDA
being a static theory, a possible way of interpreting this is that
magnetic moments undergo important dynamical fluctuations. Indeed, in Refs.~\cite{Hansmann_localmoment_prl,Toschi_moment_screening_FeSC},
Hansmann and coworkers performed LDA+DMFT calculations of the
local spin-spin correlation function $\chiloc(\tau)=\langle
S^z(0)\cdot S^z(\tau)\rangle$ in the paramagnetic phase and looked at the short-time
(high-energy) fluctuating local moment, finding that its instantaneous
value $\langle \vec{S}^2\rangle$ is rather large but rapidly decays (after
typically a few femtoseconds) due to the screening in a  metallic
environment. 
The value $m_{\mathrm{loc}}=g\mu_B[3\chiloc(\tau=0)]^{1/2}\simeq 3.68\mu_B$ was found for
LaFeAsO, with a similar value reported in
Ref.~\cite{Aichhorn_full_charge_SC_LaFeAsO} and somewhat larger in
Ref.~\cite{Yin_kinetic_frustration_allFeSC}~\footnote{The actual values
reported in these two articles correspond only to $g\sqrt{\langle S_z^2
\rangle}$ and should be multiplied by $\sqrt{3}$}.  This corresponds
(from $m_{\mathrm{loc}}^2=(g\mu_B)^2 \Seff(\Seff+1)$) to an effective
spin per iron atom $\Seff \gtrsim 1.4$. 
From neutron scattering, Liu et al. \cite{Liu_magmoments_122} report a smaller
value $\mloc \simeq 1.8 \mu_B$ ($\Seff\simeq 1/2$) for \BFA 
\footnote{$Im\chiloc(\omega)$ can be probed by neutrons and its
integral can be related to the value of the moment. However, because
neutrons only reach frequencies of order a few $100$meV's  
which is an order of magnitude too low, reduced values
of the moment can be expected from such experiments.}.
The $\chiloc^{''}(\omega)$ they find is compared to LDA+DMFT calculations and the
agreement supports the notion of a local moment formed at a high-energy,
with little influence of doping on the high-energy
spectrum. Furthermore, the maximum of $\chiloc^{\prime\prime}(\omega)$
was found to be at $\sim 200$~meV, corresponding to a fluctuation
time-scale of $\sim 20\mathrm{fs}$.
Note that this energy scale (resp. time-scale) is an order of
magnitude smaller (resp. longer) than the bare electronic bandwidth
($\simeq 4$~eV).  Indeed, a weak-coupling itinerant picture based on
an RPA calculation~\cite{Liu_magmoments_122,Toschi_moment_screening_FeSC} would yield a
time-scale about 10 times shorter and vertex corrections were found to
be crucial (see Fig.~\ref{fig:pnictides}b).  
Experimental support for the formation of a sizeable
fluctuating local moment at high-energy also stems from 
from fast spectroscopic probes
such as X-ray absorption~\cite{bondino_exchange_multiplets_prl_2008,Kroll_O_K-edge_1111} and core-level
photoemission~\cite{Vilmercati_3score_PES_strongHund}.
The importance of the Hund's coupling in properly accounting also for the
magnetic long-range order of these compounds~\footnote{The nesting
  picture has been shown to be unable to describe some key aspects of
  the SDW ordered phase, such as the difference in the magnetic
  ordering of \BFA and FeTe \cite{Johannes_local-moment}.}  has 
been emphasized by theoretical studies both from the strong
coupling~\cite{Zhou_Hund_moment,Yin_spinFermion_unified_prl,Yin_kinetic_frustration_allFeSC,Aichhorn_full_charge_SC_LaFeAsO} 
and weak coupling viewpoints~\cite{Johannes_local-moment,Raghuvanashi_Hund_stabilizes_pi-0_mag}.

%
As mentioned above, it has first been proposed in Ref.~\cite{haule09} that the Hund's coupling, besides causing moment
formation at high-energy, is also responsible for the low energy correlation effects in the metallic phase of
these compounds, hence making them `Hund's metals' (a term coined in Ref.~\cite{Yin_kinetic_frustration_allFeSC}).
This point of view has been further confirmed and elaborated upon in
several theoretical studies, mostly based on the LDA+DMFT methodology.
Aichhorn and coworkers\cite{Aichhorn_FeSe} and Liebsch and
Ishida\cite{Liebsch_FeSe_spinfreezing} found that the chalcogenide
FeSe displays local moments down to low temperature, together with
`bad metallic' behaviour characterized by a large scattering rate for some of the orbitals. 
This is a manifestation of the `spin-freezing' behaviour discussed in
Sec.~\ref{sec:spin_freezing}.  These calculations also reveal a strong
tendency to orbital differentiation (present in all materials
but more pronounced for the 
chalcogenides~\cite{Aichhorn_FeSe,Liebsch_FeSe_spinfreezing,Yin_kinetic_frustration_allFeSC, Ferber_LiFeAs_LDA_DMFT,yin_powerlaws_arxiv_2012}), 
with the \ttg-like orbitals more correlated than the \eg ones (see below).
The importance of Hund's coupling for LiFeAs was also emphasized recently in 
Ref.~\cite{Ferber_LiFeAs_LDA_DMFT}. 

The non-Fermi liquid power law regime associated with the onset of the
spin-freezing behaviour (Sec.~\ref{sec:spin_freezing}) has been
revealed very clearly in a recent study of doped \BFA
~\cite{Werner_122_dynU}. The crossover line between the Fermi-liquid
and non-Fermi liquid power-law behaviours found by these authors as a
function of doping and temperature is reproduced in
Fig.\ref{fig:pnictides}c. 
Very recently, such power-laws have been reported and discussed for chalcogenides 
as well~\cite{yin_powerlaws_arxiv_2012}. 
Based on these studies and on the general considerations presented
earlier in this article, one may want to position the different
Fe-based material on a diagram similar to the one of
Fig.~\ref{fig:Zcontour}, as a function of the filling of the d-shell
and strength of interaction, see Fig.\ref{fig:pnictides}d.
It is seen that hole-doping takes these materials deeper into the
strongly correlated spin-freezing regime and electron doping restores
a more itinerant Fermi liquid behaviour.
With this perspective in mind, some authors have recently pictured the
hole-doped materials as being in the proximity of the $d^5$ Mott
insulating state i.e as derived from this insulator by electron
doping~\cite{Liebsch_FeSe_spinfreezing,Liebsch_pseudogap,Misawa_d5-proximity_magnetic}. 
We note in this respect that the Mn-based materials, with nominal
$d^5$ composition, are indeed insulators~\cite{pandey_BaKMnAs2_prl_2012}, 
as expected from the much lower value of $U_c$ for a half-filled shell.

Insights into the qualitative difference between Mott-correlated and Hund-correlated metals have been 
obtained within LDA+DMFT by focusing on atomic histograms~\cite{Kutepov_pnictides_parameters-GW,Yin_kinetic_frustration_allFeSC}. 
These histograms register the probability of occurrence of each atomic state, resolved with respect to the atomic charge 
and the energy of the state. 
They reveal that charge fluctuations are substantial in these materials, in contrast to a metal close to a Mott transition in which 
valence fluctuations are suppressed. Here in contrast, the probability is highest for  $N=6$ and $N=7$ states, 
is still sizeable for $N=5$ and non-negligible for $N=4,N=8$ states. 
Furthermore, within a given $N$ the high-spin atomic ground state has the largest
probability~\cite{Kutepov_pnictides_parameters-GW}, but other states are also often visited, unlike in heavy fermions. 
It is also argued~\cite{Yin_kinetic_frustration_allFeSC,Kutepov_pnictides_parameters-GW} that, 
since the most probable $N=6$ and $N=7$ states span an energy range of over 6eV, this
broadens the atomic excitations (Hubbard bands) and makes them
difficult to be resolved in photoemission spectroscopy, explaining why they are actually not observed. 
These considerations highlight the itinerant nature of these systems, yet dominated by the correlation 
effects due to Hund's rule coupling.
Note that valence fluctuations on individual sites imply a corresponding change of the local 
effective interaction (Sec.~\ref{sec:mottgap},\cite{vandermarel_sawatzky_prb_1988}). 
Local aspects of this physics are fully taken into account by DMFT, 
but inter-site correlations may also play a role and require a treatment beyond single-site DMFT.  

%
%
On the experimental side, optical measurements have been interpreted as revealing the importance of the 
Hund's rule coupling. 
Besides a reduction of the Drude weight and thus of the electron kinetic energy~\cite{Qazilbash_correlations_pnictides} which testifies 
for correlations, optical measurements on \BFA\, show that 
the spectral weight $\int_0^\Omega \sigma(\omega) d \omega$ is suppressed upon cooling
down around $3000$~cm$^{-1}$, the lost spectral weight being recovered above
$8000$~cm$^{-1}$. This energy scale, first reported in Hu et al.~\cite{Hu_Optics_SDW_origin}, is interpreted as a signature of Hund's coupling~\cite{Wang_pseudogap_optics,Schafgans_optics_Co122_correlations_Hund}.

%
Soon after the discovery of high-$T_c$ superconductivity in iron pnictides it was also pointed
out~\cite{demedici_3bandOSMT,demedici_Genesis} that the general
features of the electronic structure of these materials constitute an
ideal ground for orbital-selective physics caused by electronic
correlations, and for the formation of localized magnetic moments
coexisting with metallic properties.  
An important mechanism is the role of "band-decoupler" played by the Hund's coupling, discussed in
Sec. \ref{sec:osmt}.

Indeed several theoretical studies~\cite{Aichhorn_FeSe,Laad_SusceptibilityPnictides_OSM,Craco_FeSe,Yin_kinetic_frustration_allFeSC,YuSi_LDA-SlaveSpins_LaFeAsO, yu_multiorbital_Mott_prb_2011,Ferber_LiFeAs_LDA_DMFT}
have reported strong orbital differentiation (e.g. in the degree of
coherence), in particular in the arguably more strongly correlated iron
chalcogenides, or in pnictides for correlation strengths somewhat
larger than the physical estimate\cite{Shorikov_LaFeAsO_OSMT}. In
general, $t_{2g}$ orbitals in these calculations show stronger mass
enhancements and lower coherence than the $e_g$'s.
In parallel, phenomenological models based on the coexistence of
localized and itinerant electrons were developed in order to explain the
magnetic and superconducting properties of iron pnictides~\cite{Kou_OSMT_pnictides} 
and their evolution under pressure~\cite{Hackl_Vojta_OSMT_pnictides}. 
Note however that there is not necessarily a direct connection between these 
two components (localized and itinerant) and the two types of orbitals (\ttg, \eg).  
Superexchange between well-formed local moments has been suggested as an explanation for both the collinear AF order
coexisting with metallic properties, and the linear dependence on temperature of the magnetic susceptibility in the paramagnetic phase
(although Ref.~\cite{Skornyakov_linear_susc_LaFeAsO} reproduces this behaviour, already in a purely local picture, 
due to orbital differentiation). 
Fluctuating local moments also hint at a possible pairing mechanism for superconductivity through spin fluctuations.

On the experimental side, several evidences for the coexistence of local moments and itinerant electrons have been reported. 
Inelastic neutron scattering on FeTe$_{0.35}$Se$_{0.65}$~\cite{Xu_neutrons_FeTeSe_local-itinerant} show a significant temperature independent magnetic moment (obtained by integrating the magnetic spectral weight up to 12meV), indicating that a large energy scale (i.e. states at an energy much larger than the temperature) is involved in the formation of this moment. A picture based on itinerant (albeit renormalized) electrons alone cannot explain such a magnetic response.
Analogously, nuclear magnetic resonance  data on FeSe$_{0.42}$Se$_{0.58}$ \cite{Arcon_NMR_EPR_FeSeTe_local-itinerant} show a Knight shift scaling with the local spin susceptibility measured by electron paramagnetic resonance and not with the bulk magnetic susceptibility, an evidence  interpreted as arising, in pure single crystalline samples, from intrinsically localized states coupled to quasiparticles.
ARPES measurements of the Fermi velocity in each Fermi surface sheet, in the arguably less correlated potassium-doped \BFA\, reported orbital dependent mass renormalizations~\cite{Ding_Arpes_BaK}. Accordingly, a model of two electronic fluids with different coherence properties was needed to interpret the magnetoresistance data in the cobalt-doped compound~\cite{Yuan_magnetoresistance_122_local-itinerant}.

Overall, a substantial orbital differentiation, induced by Hund's coupling, in the degree of correlation and localization of the conduction electrons 
associated with the different Fe orbitals appears to play a role in the physics of iron-based superconductors. 
To what extent and how strongly in each family of materials is still an issue for future investigation.

Finally, let us emphasize that Hund's coupling-induced correlations are relevant 
to other iron compounds~\cite{vanacker_moments_ironmaterials_prb_1988}, 
such as e.g. FeSi~\cite{arita_FeSi_prb_2008,tomczak_FeSi_PNAS_2012}.

\section{Conclusion - Future Directions}

In this article, we have emphasized that the Hund's coupling plays an essential role in the physics of 
multi-orbital materials. It induces strong electronic correlations in itinerant materials which are not 
in close proximity to a Mott insulating state. This is especially relevant to transition-metal oxides 
of the $4d$ series and to iron pnictides and chalcogenides. A global picture has recently emerged, 
which has been reviewed in this article. 

%
Some key questions remain unanswered however, to be addressed in future investigations. 
As reviewed above, the Fermi liquid scale $\TFL$ is strongly reduced by the Hund's coupling, and 
a non Fermi-liquid state with frozen local moments and power-law self-energy applies for 
$T>\TFL$. A precise theoretical understanding of this regime is still missing however. 
Is this regime associated with a specific unstable fixed point 
of the underlying effective impurity model, within a single-site DMFT approach ?  
This would yield the fascinating possibility that there is 
something universal to be learnt about the crossover between the very high-temperature regime 
of  quasi isolated atoms and the very low-temperature Fermi liquid.

Much work also remains to be done about the interplay of the effects described in this article 
with magnetic long-range order, a topic to which we have devoted only little discussion. 
How the development of inter-site magnetic correlations modifies the local picture reviewed here 
is to be addressed using other approaches, such as cluster extensions of DMFT. 

Although several indications of the key role played by the Hund's coupling have been reviewed in this article, 
a direct `smoking-gun' evidence would be invaluable. 

%
Finally, there are some  important topics that we have not covered in this article. 
These include: 
the role of Hund's coupling in stabilizing the ferromagnetic state~\cite{vanvleck_rmp_1953} in transition metals and other materials;  
the physics of negative (antiferromagnetic) Hund's coupling, which can occur due to the Jahn-Teller 
coupling and is important for the physics of fullerides~\cite{capone_rmp_fullerides_2009}; 
the possibility of Hund's coupling mediated pairing and superconductivity (see e.g.~\cite{norman_1994,han_2004}); 
the role of the Hund's coupling in heavy fermion compounds and in low-dimensional systems.  
Last but not least, the interplay of the Hund's coupling with the spin-orbit coupling is a topic of considerable current 
interest and of special relevance to the physics of transition-metal oxides of the $5d$ series.

\section*{Disclosure Statement}

The authors are not aware of any affiliations, memberships, funding, or financial holdings that might be perceived as affecting the objectivity of this review.

\section*{Acknowledgments}
We  are grateful to M.~Aichhorn, H.~Alloul, S.~Biermann, M.~Capone, M.~Casula, M.~Ferrero, A.~Fujimori, E.~Gull, P.~Hansmann, K.~Haule,
M.~Imada, G.~Kotliar, J.~Kune\v{s}, A.~Liebsch, C.~Martins, I.~Mazin, A.~J.~Millis, M.~Randeria, G.~Sangiovanni, 
G.~Sawatzky, Y.~Tokura, A.~Toth, T.~Uemura,
D.~van der Marel, L.~Vaugier, P.~Werner,  and R.\v{Z}itko for very useful discussions and remarks. 
This work was supported by the Partner University Fund, the Agence Nationale de 
la Recherche (grants ANR-09-RPDOC-019-01, ANR-2010-BLAN-040804, PNICTIDES), 
the Slovenian Research Agency (under contract J1-0747), the 
Swiss National Science Foundation MaNEP program and the JST-CREST program.  
Computer time was provided by IDRIS/GENCI under Grant 2011091393. 

\appendix
\section{Appendix: Two-orbital hamiltonian}
\label{sec:appendix}
\begin{small}

In this appendix, we provide details on the different hamiltonians relevant to the case of two orbitals. 
The orbital isospin generators read in this case (with $\vec{\tau}$ the Pauli matrices):
\begin{equation} 
\vec{T}=\frac{1}{2}\sum_\sigma d^+_{m\sigma}\vec{\tau}_{m\mp}d_{\mp\sigma}
\end{equation}
The expression of the five terms in the generalized Kanamori hamiltonian $H_{\mathrm{GK}}$, Eq.~(\ref{eq:ham_kanamori_general}), 
read in terms of charge, spin and orbital-isospin generators: 
\begin{eqnarray} \nonumber 
&\sum_m \hnmu\hnmd\,=\,
\hN^2/4+T_z^2-\hN/2\,\,\,,\,\,\,
\sum_{m\neq\mp} \hnmu\hnpmd\,=\,
-S_z^2-T_z^2+\hN/2 \\
&\sum_{m<\mp,\sigma} \hn_{m\sigma}\hn_{\mp\sigma}\,=\,
\hN^2/4+S_z^2-\hN/2 \\
&\sum_{m\neq\mp} d^+_{m\spinup}d_{m\spindown}\,d^+_{\mp\spindown}d_{\mp\spinup}\,=\,
(\vec{S}^2-\vec{T}^2)/2+T_z^2-S_z^2 \,\,\,,\,\,\,
\sum_{m\neq\mp} d^+_{m\spinup}d^+_{m\spindown}\,d_{\mp\spindown}d_{\mp\spinup} \,=\,T_x^2-T_y^2  \nonumber
\end{eqnarray}
Note also the relation: 
\begin{equation}
(\hN-2)^2+2\vec{S}^2+2\vec{T}^2=4
\end{equation}
As for \ttg, the Kanamori hamiltonian (\ref{eq:ham_kanamori}) is exact for an \eg doublet, but 
in this case cubic symmetry itself implies that $\Up=U-2J$~\cite{sugano_multiplets_book}. 
The \eg Kanamori hamiltonian can be written as:
\begin{equation}
H_{e_g} = (U-J)\frac{\hN(\hN-1)}{2} + 2J(T_x^2+T_z^2)  -J \hN
\label{eq:ham_eg}
\end{equation}
It is seen that no continuous orbital symmetry remains, due to the total quenching of orbital 
angular momentum for an \eg doublet. For a spherically symmetric atom, $U$ and $J$ can again be related to 
Racah-Slater parameters, as~\cite{sugano_multiplets_book} :
\begin{equation}
U=\Up+2J=F^0+\frac{4}{49}F^2+\frac{4}{49}F^4=A+4B+3C\,\,\,,\,\,\,
J=\frac{3}{49}F^2+\frac{5}{147}F^4=4B+C
\label{eq:slater_eg}
\end{equation}
The generalized Kanamori hamiltonian (\ref{eq:ham_kanamori_general}) can also be written in terms of the 
different generators as: 
\begin{equation}
\frac{1}{4}(U+\Up-J+J_X)(\hN-2)^2+J_X\vec{T}^2+(U-\Up-J_X)T_z^2+(J_X-J)S_z^2+J_P(T_x^2-T_y^2)
\end{equation}
in which we have focused on the particle-hole symmetric case, hence omitting a term $\hN(U+2\Up-J)/2$.
%
%
Two special cases are worth mentioning, for future reference:
\begin{itemize}
\item Full spin and orbital invariance $U(1)_C\otimes SU(2)_S\otimes SU(2)_O$ is realized for $J_P=0$, $J_X=J$ and 
$\Up=U-J$ (note: not $\Up=U-2J$). This actually applies to an arbitrary number of orbitals, and yields the  
hamiltonian Eq.~(\ref{eq:ham_DN}) introduced by Dworin and Narath~\cite{dworin_prl_1970} 
in the context of magnetic impurities:
\begin{equation}
\frac{1}{2}(U-\frac{J}{2})(\hN-2)^2+J\vec{T}^2=\frac{1}{2}(U-\frac{3J}{2})(\hN-2)^2-J\vec{S}^2+2J
\label{eq:ham_dworin_2orbital}
\end{equation}
 \item Setting $J_P=0$, $J_X=J$ and $\Up=U$, we obtain a hamiltonian which still implements the essence of Hund's rule 
 physics while maintaining a partial $O(2)$ orbital symmetry (it commutes with $\vec{T}^2$ and $T_z$). 
 This hamiltonian was introduced by Caroli, Lederer and Saint-James~\cite{caroli_hund_prl_1969} 
 (see also \cite{deleo_2orbital_impurity_prb_2004,leo04thesis}) and reads: 
\begin{equation}
\frac{U}{2} (\hN-2)^2 + J(\vec{T}^2-T_z^2)
\end{equation} 
\end{itemize}

\end{small}

\end{document}